\documentclass[prd, twocolumn, nofootinbib, letterpaper, showpacs]{revtex4}

\usepackage{amsmath}
\usepackage{amssymb}
\usepackage{graphicx}

\newcommand{\dif}{\mathrm{d}}
\newcommand{\form}[1]{\boldsymbol{\mathrm{#1}}}
\newcommand{\switch}{[1 \leftrightarrow 2]}
\newcommand{\pder}[2]{\partial #1 / \partial #2}
\newcommand{\fpder}[2]{\frac{\partial #1}{\partial #2}}
\newcommand{\fpdert}[2]{\frac{\partial^2 #1}{\partial #2^2}}
\newcommand{\fpdertm}[1]{\frac{\partial^2 #1}{\partial r \partial t}}

\newlength{\figwidth}
\begin{document}
\addtolength{\figwidth}{0.97\columnwidth}

\title{Stability of spherically symmetric solutions in modified
  theories of gravity}


\author{Michael D.\ Seifert}
\affiliation{Dept.\ of Physics, University of Chicago, 5640 S. Ellis
  Ave., Chicago, IL, 60637}
\email{seifert@uchicago.edu}

\begin{abstract}
In recent years, a number of alternative theories of gravity have been 
proposed as possible resolutions of certain cosmological problems or
as toy models for possible but heretofore unobserved effects.
However, the implications of such theories for the stability
structures such as stars have not been fully investigated.  We use our
``generalized variational principle'', described in a previous work
\cite{GVP}, to analyze the stability of static spherically symmetric
solutions to spherically symmetric perturbations in three such
alternative theories: Carroll \textsl{et al.}'s $f(R)$ gravity,
Jacobson \& Mattingly's ``Einstein-{\ae}ther theory'', and
Bekenstein's TeVeS.  We find that in the presence of matter, $f(R)$
gravity is highly unstable;  that the stability conditions for
spherically symmetric curved vacuum Einstein-{\ae}ther backgrounds are
the same as those for linearized stability about flat spacetime, with
one exceptional case; and that the ``kinetic terms'' of vacuum TeVeS
are indefinite in a curved background, leading to an instability. 
\end{abstract}

\maketitle

\section{Introduction}

The idea of using a variational principle to bound the spectrum of an
operator is familiar to anyone who has taken an undergraduate course
in quantum mechanics:  given a Hamiltonian $H$ whose spectrum is
bounded below by $E_0$, we must have for any normalized state $| \psi
\rangle$ in our Hilbert space $E_0 \leq \langle \psi | H | \psi \rangle$.
We can thus obtain an upper bound on the ground state energy of the
system by plugging in various ``test functions'' $|\psi \rangle$,
allowing these to vary, and finding out how low we can make the
expectation value of $H$.  

This technique is not particular to quantum
mechanics;  rather, it is a statement about the properties of
self-adjoint operators on a Hilbert space.  In particular, we can make
a similar statement in the context of linear field theories.  Suppose
we have a second-order linear field theory, dependent on some
background fields and some dynamical fields $\psi^\alpha$, whose
equations of motion can be put into the form 
\begin{equation}
  \label{simpleform}
  - \frac{\partial^2}{\partial t^2} \psi^\alpha =
    {\mathcal{T}^\alpha}_\beta \, \psi^\beta  
\end{equation}
for some linear spatial differential operator
$\mathcal{T}$.  Suppose, further, that we can find an inner product
$(\cdot, \cdot)$ such that $\mathcal{T}$ is self-adjoint under this
inner product, i.e., 
\begin{equation}
  \label{simpleTsym}
  (\psi_1, \mathcal{T} \psi_2) = (\mathcal{T} \psi_1, \psi_2)
\end{equation}
for all $\psi_1$ and $\psi_2$.  Then the spectrum of $\mathcal{T}$
will correspond to the squares of the frequencies of oscillation of
the system;  and we can obtain information about the lower bound of
this spectrum via a variational principle of the form
\begin{equation}
  \label{basicvarprin}
  \omega_0^2 \leq \frac{ (\psi, \mathcal{T} \psi)}{(\psi, \psi)}.
\end{equation}
Moreover, if we find that we can make this quantity negative for some
test function $\psi$, then the system is unstable
\cite{Waldstability};  and we can obtain an upper bound on the
timescale of this instability by finding how negative $\omega_0^2$
must be. 

One might think that this would be a simple way to analyze any
linear (or linearized) field theory;  in practice, however, it is
easier said than done.  A serious problem arises when we consider
linearizations of covariant field theories.  In such theories, the
linearized equations of motion are not all of the form
\eqref{simpleform};  instead, we obtain non-deterministic equations
(due to gauge freedom) and equations which are not evolution equations
but instead relate the initial data for the fields $\psi^\alpha$
(since covariant field theories on a spacetime become constrained when
decomposed into ``space + time''.)  Further, even if we have equations
of the form \eqref{simpleform}, it is not always evident how to find
an inner product under which the time-evolution operator $\mathcal{T}$
is self-adjoint (or even if such an inner product exists.) Thus, it
would seem that this method is of limited utility in the context of
linearized field theories.  

In spite of these difficulties, Chandrasekhar successfully derived a
variational principle to analyse the stability of spherically
symmetric solutions of Einstein gravity with perfect-fluid sources
\cite{Chandra1, Chandra2}.  The methods used in these works to
eliminate the constraint equations, put the equations in the form
\eqref{simpleform}, and obtain an inner product seemed rather
particular to the theory he was examining, and it was far from certain
that they would generalize to an arbitrary field theory.  In a recent
paper \cite{GVP}, we showed that these methods did in a certain sense
``have to work out'', by describing a straightforward procedure by
which the gauge could be fixed, the constraints could be solved, and
an inner product could be obtained under which the resulting
time-evolution operator $\mathcal{T}$ is self-adjoint.  We review this
procedure in Section \ref{GVPsec} of the paper.  We then use this
``generalized variational principle'' to analyze three alternative
theories of gravity which have garnered some attention in recent
years:  $f(R)$ gravity, the current interest in which was primarily
inspired by the work of Carroll \textsl{et al.~}\cite{CDTT};
Einstein-{\ae}ther theory, a toy model of Lorentz-symmetry breaking
proposed by Jacobson and Mattingly \cite{Aether1, Aetherwave};  and
TeVeS, a covariant theory of MOND proposed by Bekenstein \cite{TeVeS}.
We apply our techniques to these theories in Sections \ref{fRsec},
\ref{aethersec}, and \ref{TeVeSsec}, respectively.  
 
We will use the sign conventions of \cite{WaldGR} throughout.  Units
will be those in which $c = G = 1$.  

\section{Review of the Generalized Variational Principle \label{GVPsec}} 

\subsection{Symplectic dynamics}

We first introduce some necessary concepts and notation.  Consider a
covariant field theory with an action of the form 
\begin{equation}
  S = \int \form{\mathcal{L}} = \int \mathcal{L}[\Psi] \form{\epsilon}
\end{equation}
where $\mathcal{L}[\Psi]$ is a scalar depending on some set
$\Psi$ of dynamical tensor fields including the spacetime metric.
(For convenience, we will describe the gravitational degrees of freedom
using the inverse metric $g^{ab}$ rather than the metric $g_{ab}$
itself.)  To obtain the equations of motion for the dynamical fields,
we take the variation of the four-form $\mathcal{L} \form{\epsilon}$ with
respect to the dynamical fields $\Psi$:
\begin{equation}
  \label{genLvar}
  \delta \left(\mathcal{L} \form{\epsilon} \right) = \left(
  \mathcal{E}_\Psi \, 
  \delta \Psi + \nabla_a \theta^a [\Psi, \delta \Psi] \right)
  \form{\epsilon} 
\end{equation}
where a sum over all fields comprising $\Psi$ is implicit in the first
term.  Requiring that $\delta S = 0$ under this variation then implies
that the quantities $\mathcal{E}_\Psi$ vanish for each dynamical field
$\Psi$.  
 
The second term in \eqref{genLvar} defines the vector field $\theta^a
[\Psi, \delta \Psi]$.  The three-form $\form{\theta}$ dual to this
vector field (i.e., $\theta_{bcd} = \theta^a \epsilon_{abcd}$) is the
``symplectic potential current.''  Taking
the antisymmetrized second variation of this quantity, we then obtain
the symplectic current three-form $\form{\omega}$ for the theory:
\begin{equation}
  \label{genomega}
  \form{\omega}[\Psi; \delta_1 \Psi, \delta_2 \Psi] = \delta_1
  \form{\theta}[\Psi, \delta_2 \Psi] - \delta_2 \form{\theta}[\Psi,
  \delta_1 \Psi].
\end{equation}
In terms of the vector field $\omega^a$ dual to $\form{\omega}$, this
can also be written as
\begin{equation}
  \label{genomegaalt}
  \omega^a \epsilon_{abcd} = \delta_1 (\theta_2^a \epsilon_{abcd} ) -
  \delta_2 ( \theta_1^a \epsilon_{abcd} ).
\end{equation} 
The symplectic form for the theory is then obtained by integrating
the pullback of this three-form over a spacelike three-surface
$\Sigma$: 
\begin{equation}
  \label{gensympform}
  \Omega[\Psi; \delta_1 \Psi, \delta_2 \Psi] = \int_\Sigma
  \form{\bar{\omega}}[\Psi; \delta_1 \Psi, \delta_2 \Psi]. 
\end{equation}
If we define $n^a$ as the future-directed timelike normal to $\Sigma$
and $\boldsymbol{e}$ to be the induced volume three-form on $\Sigma$
(i.e., $e_{bcd} = n^a \epsilon_{abcd}$), this can be written in terms
of $\omega^a$ instead:
\begin{equation}
  \label{gensympformalt}
  \Omega = - \int_\Sigma (\omega^a n_a) \boldsymbol{e}.
\end{equation}
In performing the calculations which follow, it is this second
expression for $\Omega$ which will be most useful to us.

\subsection{Obtaining a variational principle}

In \cite{GVP}, we presented a procedure by which a variational
principle for spherically symmetric perturbations of static,
spherically symmetric spacetimes could generally be obtained.  For our
purposes, we will outline this method;  further details can be found
in the original paper.

The method described in \cite{GVP} consists of the following steps:
\begin{enumerate}
\item Vary the action to obtain the equations of motion
  $(\mathcal{E}_G)_{ab}$ corresponding to the variation of the metric,
  as well as any other equations of motion $\mathcal{E}_A$
  corresponding to the variations of any matter fields present.  This
  variation will also yield the dual $\theta^a$ of the symplectic
  current potential;  take the antisymmetrized variation of this
  quantity (as in \eqref{genomegaalt}) to obtain the symplectic form
  \eqref{gensympformalt}.

\item Fix the gauge for the metric, and choose an appropriate set of
  spacetime functions to describe the matter fields.  Throughout this
  paper, we will choose our coordinates such that the metric takes the
  form 
  \begin{equation}
    \label{spheremet}
    \dif s^2 = -e^{2 \Phi(r,t)} \, \dif t^2 + e^{2 \Lambda(r,t)} \, \dif
    r^2 + r^2 \dif \Omega^2
  \end{equation}
  for some functions $\Phi$ and $\Lambda$.\footnote{Such a set of
    coordinates can always be found for a spherically symmetric
    spacetime \cite{MTW}.} 
  Our spacetimes will be static at zero-order (i.e., $(\pder{}{t})^a$
  is a Killing vector field at zero-order), but non-static in
  first-order perturbations.  We will  therefore have $\Phi(r,t) =
  \Phi(r) + \phi(r,t)$, where $\phi$ is a first-order quantity;
  similarly, we define the first-order quantity $\lambda$ such that
  $\Lambda(r,t) = \Lambda(r) + \lambda(r,t)$.  In other words,
  \begin{equation}
    \delta g^{tt} = 2 e^{-2 \Phi(r)} \phi(r,t)
  \end{equation}
  and
  \begin{equation}
    \delta g^{rr} = -2 e^{-2 \Lambda(r)} \lambda (r,t).
  \end{equation} 
  Similarly, all matter fields will be static in the background, but
  possibly time-dependent at first order.  

\item Write the linearized equations of motion and the symplectic
  form in terms of these perturbational fields (metric and matter.) 

\item Solve the linearized constraints.  One of the main results of
  \cite{GVP} was to show that this can be done quite generally for
  spherically symmetric perturbations off of a static, spherically
  symmetric background.  Specifically, suppose the field content
  $\Psi$ of our theory consists of the inverse metric $g^{ab}$ and a
  single tensor field $A^{a_1 \dots a_n} {}_{b_1 \dots b_m}$.  (The
  generalization to multiple tensor fields is straightforward.)  Let
  $(\mathcal{E}_A)_{a_1 \dots a_n} {}^{b_1 \dots b_m}$ denote the
  equation of motion associated with $A^{a_1 \dots a_n} {}_{b_1 \dots 
  b_m}$. We define the \emph{constraint tensor} as
  \begin{multline}
    \label{Cdef}
    C_{cd} = 2 (\mathcal{E}_G)_{cd} \\ 
    - g_{ce} \sum_i A^{a_1 \dots a_n} {}_{b_1 \dots d \dots b_m}
    (\mathcal{E}_A)_{a_1 \dots a_n} {}^{b_1 \dots e \dots b_m} \\ 
    + g_{ce} \sum_i A^{a_1 \dots e \dots a_n} {}_{b_1 \dots b_m}
    (\mathcal{E}_A)_{a_1 \dots d \dots a_n} {}^{b_1 \dots b_m}  
  \end{multline}
  where the summations run over all possible index ``slots'', from $1$
  to $n$ and from $1$ to $m$ for the first and second summation
  respectively.  It can then be shown that if the background equations
  of motion hold, and the matter equation of motion also holds to
  first order, the perturbations of the tensor $C_{ab}$ will satisfy
  \begin{equation}
    \label{Fdef1}
    \fpder{F}{t} = - r^2 e^{\Phi - \Lambda} \delta C_{rt}
  \end{equation}
  and
  \begin{equation}
    \label{Fdef2}
    \fpder{F}{r} = - r^2 e^{\Lambda - \Phi} \delta C_{tt}
  \end{equation}
  for some quantity $F$ which is linear in the first-order fields.
  Moreover, the first-order constraint equations $\delta C_{tt} = 
  \delta C_{tr} = 0$ will be  satisfied if and only if $F = 0$.  We
  can then solve this equation  algebraically for one of our
  perturbational fields, usually the metric perturbation $\lambda$.
  We will refer to the equation $F = 0$ as the \emph{preconstraint
  equation}. 

\item  Eliminate the metric perturbation $\phi$ from the equations.
  As $\phi$ cannot appear without a radial derivative (due to
  residual gauge freedom)\footnote{The
    exception to this statement is when the matter fields under
    consideration have non-zero components in the $t$-direction.  In
    this case, $\phi$ and its derivatives can appear in linear
    combination with perturbations of the matter fields;  however, a
    redefinition of the matter fields will suffice to eliminate such
    cases. \label{phifoot}}, we must find an algebraic equation for
  $\pder{\phi}{r}$.  The first-order equation $\delta C_{rr} = 0$,
  will, in general, serve this purpose \cite{GVP}. 
  Use the above relations for $\lambda$ and $\pder{\phi}{r}$ to
  eliminate the metric perturbations completely from the
  perturbational equations of motion and the symplectic form,
  leaving a ``reduced'' set of equations of motion and a ``reduced''
  symplectic form solely in terms of the matter variables.
  
\item Determine if the reduced equations of motion take the form
  \eqref{simpleform}.  If so, read off the time-evolution operator
  $\mathcal{T}$.
  
\item Determine if the symplectic form, written in terms of the
  reduced dynamical variables $\psi^\alpha$, is of the form 
  \begin{equation}
    \label{Wdef}
    \Omega(\Psi;  \psi_1^\alpha, \psi_2^\alpha) = \int_\Sigma
    \form{W}_{\alpha \beta} \left( \fpder{\psi_1^\alpha}{t} \psi_2^\beta
    - \fpder{\psi_2^\alpha}{t} \psi_1^\beta \right)
  \end{equation}
  for some three-form $\form{W}_{\alpha \beta}$.  Then (as we showed
  in \cite{GVP}) we must have 
  $\form{W}_{\alpha \beta} = \form{W}_{\beta \alpha}$ and
  \begin{equation}
    \label{Tsym}
    \int_\Sigma \form{W}_{\alpha \beta} \psi_1^\alpha \mathcal{T}^\beta
	{}_\gamma \psi_2^\gamma  = \int_\Sigma \form{W}_{\alpha \beta}
	\psi_2^\alpha \mathcal{T}^\beta {}_\gamma \psi_1^\gamma.
  \end{equation}
  
\item  Determine whether $\form{W}_{\alpha \beta}$ is positive
  definite, in the sense that
  \begin{equation}
    \label{Wposdef}
    \int_\Sigma \form{W}_{\alpha \beta} \psi^\alpha \psi^\beta \geq 0
  \end{equation}
  for all $\psi^\alpha$, with equality holding only when $\psi^\alpha = 
  0$.\footnote{If $\form{W}_{\alpha \beta}$ is \emph{negative}
    definite in this sense, we use the negative of this quantity as
    our inner product, and the construction proceeds identically.} If
  this is the case, we can define an inner product on the space of all
  reduced fields: 
  \begin{equation}
    \label{geninnerprod}
    (\psi_1, \psi_2) \equiv \int_\Sigma \form{W}_{\alpha \beta}
    \psi_1^\alpha \psi_2^\beta.
  \end{equation}
  Equation \eqref{Tsym} then shows that $\mathcal{T}$ is a symmetric
  operator under this inner product.  Thus, we can write down our
  variational principle of the form \eqref{basicvarprin}.
\end{enumerate}

It is important to note that each step of this procedure is clearly
delineated.  While the procedure can fail at certain steps (the
reduced equations of motion can fail to be of the form
\eqref{simpleform}, for example, or $\form{W}_{\alpha \beta}$ can fail
to be positive definite), there is not any ``art'' required to apply
this procedure to an arbitrary covariant field theory.  (In practice,
as we shall see, there are certain shortcuts that may arise which we
can exploit; however, the ``long way'' we have described here will
still work.)  In the following three sections, we will use this
formalism to analyze the stability of $f(R)$ gravity,
Einstein-{\ae}ther theory, and TeVeS. 

\section{$f(R)$ gravity \label{fRsec}}

\subsection{Theory}

In $f(R)$ gravity, the Ricci scalar $R$ in the Einstein-Hilbert action
is replaced by an arbitrary function of $R$, leaving the rest of the
action unchanged;  in other words, the Lagrangian four-form
$\form{\mathcal{L}}$ is of the form
\begin{equation}
\label{fRlag}
\form{\mathcal{L}} = \frac{1}{16 \pi} f(R) \form{\epsilon} +
\form{\mathcal{L}}_{\text{mat}}[A, g^{ab}]
\end{equation}
where $A$ denotes the collection of matter fields, with tensor indices
suppressed.  Taking the variation of this action with respect to the
metric, we obtain the equation of motion 
\begin{multline}
  \label{4thordereqn}
  f'(R) R_{ab} - \frac{1}{2} f(R) g_{ab} \\ - \nabla_a \nabla_b f'(R) +
  g_{ab} \Box f'(R) = 8 \pi T_{ab}
\end{multline}
where $T_{ab}$, given by
\begin{equation}
  \label{Tabdef}
  \delta \form{\mathcal{L}}_{\text{mat}} = - \frac{1}{2} (T_{ab}
  \delta g^{ab}) \form{\epsilon},
\end{equation}
is the matter stress-energy tensor, and $f'(R) = \dif f/\dif R$.

This equation is fourth-order in the metric, and as such is somewhat
difficult to deal with.  We can reduce this fourth-order equation to
two second-order equations using an equivalent scalar-tensor theory
\cite{Odintsov, Chiba}.\footnote{This equivalence was also used in the
  special case $f(R) = R - 2 \Lambda + \alpha R^2$ in \cite{Whitt}.} 
This equivalent theory contains two dynamical
gravitational variables, the inverse metric $g^{ab}$ and a scalar field
$\alpha$, in addition to the matter fields.  The Lagrangian is given by 
\begin{equation}
  \label{sctensaction}
  \form{\mathcal{L}} = \frac{1}{16 \pi} \left( f'(\alpha) R + f(\alpha) -
  \alpha f'(\alpha) \right) \form{\epsilon} +
  \form{\mathcal{L}_{\text{mat}}}[A, g^{ab}].
\end{equation}
Varying the gravitational part of action with respect to $g^{ab}$ and
$\alpha$ gives us
\begin{equation}
  \label{scvariation}
  \delta \form{\mathcal{L}} = \left(
  (\mathcal{E}_G)_{ab} \delta g^{ab} + \mathcal{E}_\alpha \delta
  \alpha + \mathcal{E}_A \delta A + \nabla_a \theta^a \right)
  \form{\epsilon} 
\end{equation}
where $\mathcal{E}_A$ denotes the matter equations of motion,
\begin{multline}
  \label{fReineq}
  (\mathcal{E}_G)_{ab} = \frac{1}{16 \pi} \bigg[ f'(\alpha) G_{ab} -
    \nabla_a \nabla_b 
    f'(\alpha) + g_{ab} \Box  f'(\alpha) \\ \left. - \frac{1}{2} g_{ab} (
    f(\alpha) - \alpha f'(\alpha)) - 8 \pi T_{ab} \right]
\end{multline}
and
\begin{equation} 
  \mathcal{E}_\alpha = f''(\alpha) (R - \alpha),
\end{equation}
and the vector $\theta^a$ is our symplectic potential current:
\begin{multline}
  \label{fRtheta}
  \theta^a = f'(\alpha)
  \theta^a_{\text{Ein}} + \theta^a_\text{mat} \\ + \frac{1}{16 \pi}
  \left( (\nabla_b 
  f'(\alpha)) \delta g^{ab} - (\nabla^a  f'(\alpha)) g_{bc} \delta
  g^{bc} \right).
\end{multline}
The vector $\theta^a_\text{mat}$ above is the symplectic potential
current resulting from variation of $\form{\mathcal{L}}_\text{mat}$,
and $\theta^a_\text{Ein}$ is the symplectic potential current for pure
Einstein gravity, i.e.,
\begin{equation}
  \label{thetaein}
  \theta^a_\text{Ein} = \frac{1}{16 \pi} \left( g_{bc} \nabla^a \delta
  g^{bc} - \nabla_b \delta g^{ab} \right).
\end{equation}
The equations of motion are then given by $(\mathcal{E}_G)_{ab} = 0$
and $\mathcal{E}_\alpha = 0$.
Assuming that $f''(\alpha) \neq 0$, this second equation implies that
$R = \alpha$, and 
substituting this relation into \eqref{fReineq} yields the equation
of motion obtained in \eqref{4thordereqn}.  Hereafter, we will use
the form of the equations obtained from the action
\eqref{sctensaction}.

\subsection{Obtaining a variational principle}

Before we examine the stability of spherically symmetric static
solutions in $f(R)$ gravity with perfect fluid matter, we must first
consider the question of whether physically realistic solutions exist.
In particular, do there exist solutions to the equations of motion
which reproduce Newtonian gravity, up to small relativistic
corrections?  There has recently been a good deal of debate on this
subject \cite{fRint1, fRint2, fRint3, fRrefute, fRsolar}.  We will not
comment directly on this controversy here except to say that if $f(R)$
gravity (or any theory) does not allow interior solutions with $R
\approx 8 \pi G \rho$ to be matched to exterior solutions with $R$
close to zero, then it is difficult to see how such a theory could
reproduce Newtonian dynamics in a nearly-flat spacetime. We will
therefore give the theory the ``benefit of the doubt,'' and assume
that such solutions exist.

An initial attempt to address the stability of such solutions was made
by Dolgov and Kawasaki \cite{DolKaw};  their perturbation analysis
implied that stars would be extremely unstable in Carroll \textsl{et
al.}'s $f(R)$ gravity, with a characteristic time scale of
approximately $10^{-26}$ seconds.  Their results, while suggestive,
nevertheless failed to take into account the constrained nature of the
theory:  the stress-energy tensor, the metric, and the scalar $\alpha$
cannot all be varied independently.  The imposition of a constraint
can, of course, change whether or not a given system is stable;  as a
trivial example, consider a particle moving in the potential $V(x,y) =
x^2 - y^2$, with and without the constraint $y =
\text{const}$.\footnote{In certain scalar-tensor theories
  \cite{Harada}, the scalar perturbations decouple from the metric and
  matter perturbations; it is then legitimate in such theories to
  ``ignore'' the constraints if we concern
  ourselves only with the scalar field.  However, this decoupling does
  not occur in $f(R)$ gravity theory.  To see this, we can rewrite the
  Lagrangian in terms of a rescaled metric $\tilde{g}^{ab} =
  \Omega^{-2}(\sigma) g^{ab} = e^{\sigma/2 \sqrt{3 \pi}} g^{ab}$,
  putting it in the form \cite{Odintsov, Chiba}
  \begin{equation} 
    \form{\mathcal{L}} = \left( \frac{1}{16 \pi} \tilde{R} - \frac{1}{2} 
    \nabla_a \sigma \nabla^a \sigma - V(\sigma) \right)
    \form{\tilde{\epsilon}} + \form{\mathcal{L}_\text{mat}}[A, e^{\sigma
	/ 2 \sqrt{3 \pi}} \tilde{g}^{ab}]
  \end{equation}
  where $\sigma$ is related to the scalar $\sigma$ by $f'(\alpha) = e^{
  \sigma/2 \sqrt{3 \pi}}$ and the exact form of the potential
  $V(\sigma)$ is determined by the function $f(R)$.  The decoupling of
  the scalar perturbations, however, requires $V(\sigma_0) = 0$ and
  $\Omega'(\sigma_0)/\Omega = 0$, where $\sigma_0$ is the background
  value of $\sigma$.  The former condition will not hold for a generic
  $f(R)$, and the latter condition will not hold for any $f(R)$ for which $f''(R) \neq 0$.
}
In what follows, we will show that Dolgov and Kawasaki's conclusion is,
nonetheless, correct:  stars in CDTT $f(R)$ gravity do in fact have an
ultra-short timescale instability. 

To describe the fluid matter, we will use the ``Lagrangian
coordinate'' formalism, as in Section V of \cite{GVP}.  In this
formalism, the fluid is described by considering the manifold
$\mathcal{M}$ of all fluid worldlines in the spacetime, equipped with
a volume three-form $\form{N}$.  If we introduce three ``fluid
coordinates'' $X^A$ on $\mathcal{M}$, with $A$ running from one to
three, then the motion of the fluid in our spacetime manifold $M$ is
completely described by a map $\chi: M \to \mathcal{M}$ associating
with every spacetime event $x$ the fluid worldline $X^A(x)$ passing
through it.  The matter Lagrangian is then given by
\begin{equation}
  \form{\mathcal{L}} = - \varrho(\nu) \form{\epsilon}
\end{equation}
where $\nu$, the ``number density'' of the fluid, is given by
\begin{equation}
  \nu^2 = \frac{1}{6} N_{abc} N^{abc}.
\end{equation}
In turn, $N_{abc}$, the ``number current'' of the fluid, is given by
\begin{equation}
  N_{abc} = N_{ABC}(X) \nabla_a X^A \nabla_b X^B \nabla_c X^C.
\end{equation}
We will be purely concerned with spherically symmetric solutions and
radial perturbations of the fluid;  thus, we will take our Lagrangian
coordinates to be of the form
\begin{align}
  X^R &= r, & X^\Theta &= \theta, & X^\Phi = \varphi
\end{align} 
in the background, and consider only perturbations $\delta X^R \equiv
\xi(r,t)$ at first order (i.e., $\delta X^\Theta = \delta X^\Phi = 0$.)

We can then easily obtain the background equations of motion;  these
are
\begin{subequations}
\begin{multline} \label{fRGtteq0}
  (\mathcal{E}_G)_{tt} = e^{2 \Phi - 2 \Lambda} \left[ - \frac{\partial^{2}}{\partial r^{2}}
  f'(\alpha) +  \left( \fpder{\Lambda}{r} - \frac{2}{r} \right)
  \frac{\partial}{\partial r} f'(\alpha) \right. \\
  \left. + \left(\frac{2}{r}
  \fpder{\Lambda}{r} + \frac{1}{r^2} (e^{2 \Lambda } - 1) \right)
  f'(\alpha) \right] \\ 
  - \frac{1}{2} e^{2 \Phi } (\alpha f'(\alpha) - f(\alpha)) - e^{2 \Phi } \varrho = 0,
\end{multline}
\begin{multline} \label{fRGrreq0}
  (\mathcal{E}_G)_{rr} = \left(\fpder{\Phi}{r} + \frac{2}{r} \right)
  \frac{\partial}{\partial  r} f'(\alpha) \\ + \left(\frac{2}{r}
  \fpder{\Phi}{r} - \frac{1}{r^2} (e^{2 \Lambda} - 1) \right) f'(\alpha)
  \\ + \frac{1}{2} e^{2 \Lambda} (\alpha f'(\alpha) - f(\alpha)) - e^{2
    \Lambda} \left( \varrho' \nu - \varrho \right) = 0,
\end{multline}
\begin{multline}
  \label{fRGththeq0}
  (\mathcal{E}_G)_{\theta \theta} = \\ 
  r^2 e^{-2 \Lambda} \left[ \frac{\partial^{2}}{\partial r^{2}} f'(\alpha) + \left( \fpder{\Phi}{r} - \fpder{\Lambda}{r} 
  + \frac{1}{r} \right) \frac{\partial}{\partial r} f'(\alpha) \right. \\ 
  \left. + \left(\fpdert{\Phi}{r} -
  \fpder{\Phi}{r} \fpder{\Lambda}{r} + \left(\fpder{\Phi}{r} \right)^2 +
  \frac{1}{r} \left( \fpder{\Phi}{r} - \fpder{\Lambda}{r} \right)
  \right)f'(\alpha) \right] \\
  - \frac{r^2}{2} (\alpha f'(\alpha) - f(\alpha))  - r^2 \left( \varrho' \nu -
  \varrho \right) = 0,
\end{multline}
\begin{multline} \label{fRscalareq}
  \mathcal{E}_\alpha = 2 f''(\alpha) e^{-2 \Lambda} \left[ -\fpdert{\Phi}{r} + \fpder{\Phi}{r}
  \fpder{\Lambda}{r} - \left(\fpder{\Phi}{r} \right)^{2} \right. \\ 
  \left. + \frac{2}{r}
  \left(\fpder{\Lambda}{r} -  \fpder{\Phi}{r} \right) + \frac{1}{r^{2}} \left( e^{2 \Lambda} - 
  1 \right)\right]  - f''(\alpha) \alpha = 0,
\end{multline}
and
\begin{equation} \label{fRhydroeq}
  (\mathcal{E}_X)_R = \varrho'' \fpder{\nu}{r} + \fpder{\Phi}{r}
  \varrho' \nu = 0,
\end{equation}
\end{subequations}
where $\varrho' = \dif \varrho/\dif \nu$.  These equations are not all
independent;  in particular, the Bianchi identity implies that
\eqref{fRGththeq0} is automatically satisfied if the other four
equations are satisfied.  Note that under the substitutions $\varrho
\to \rho$, $\varrho' \nu - \varrho \to P$, the ``matter terms'' in
these equations take on their familiar forms for a perfect fluid.

We next obtain the symplectic form for the theory.  The contribution
$\theta^a_\text{grav}$ to the symplectic potential current from the
gravitational portion of the action is given by \eqref{fRtheta}; to
obtain the symplectic form, we then take the antisymmetrized second
variation of $\form{\theta}$, as in \eqref{genomegaalt}. Performing
this variation, we find that
\begin{multline}
  \label{fRtensomega}
  \omega^a_\text{grav} = f'(\alpha) \omega^a_{\text{Ein}} + \frac{1}{2}
  \delta_1 g^{bc} \delta_2 g^{ad} g_{bc} \nabla_d (f'(\alpha)) \\ +
  \left[ \delta_1 (f'(\alpha)) \nabla_b \delta_2 g^{cd} -  \nabla_b
  \left(\delta_1 (f'(\alpha)) \right) \delta_2 g^{cd} \right] \\ \times
  (g^{ab} g_{cd} - \delta^a {}_c \delta^b {}_d )  - \switch
\end{multline}
where $\omega^a_\text{Ein}$ is defined, analogously to
$\theta^a_\text{Ein}$, to be the symplectic current associated with
pure Einstein gravity.  This is equal to \cite{BurnettWald}
\begin{equation}
  \label{omegaein}
  \omega_\text{Ein}^a = S^{a} {}_{bc} {}^d {}_{ef} (\delta_2 g^{bc}
  \nabla_d \delta_1 g^{ef} - \delta_1 g^{bc} \nabla_d \delta_2 g^{ef}),
\end{equation}
where
\begin{multline}
  S^{a} {}_{bc} {}^d {}_{ef} = \frac{1}{16 \pi} \left(  \delta^a {}_e
  \delta^d {}_c g_{bf}  - \frac{1}{2} g^{ad} g_{be} g_{cf} \right.  \\
  \left. - \frac{1}{2} \delta^a {}_b \delta^d {}_c g_{ef}  - \frac{1}{2}
  \delta^a {}_e \delta^d {}_f g_{bc}  + \frac{1}{2} g^{ad} g_{bc} g_{ef}
  \right).
\end{multline}

The contribution to the symplectic current from the matter terms in
the Lagrangian coordinate formalism was calculated in \cite{GVP};  we
simply cite the result for the $t$-component of $\omega^a$ in a static
background here:
\begin{multline}
  \label{csympcurr}
  t_a \omega^a_\text{mat} = - t_a \frac{\varrho'}{2 \nu} N_{ABC}
  \nabla_b X^B \nabla_c X^C \left[ \delta_1 g^{ad} \delta_2 X^A N_{d}
    {}^{bc} \right. \\ 
    \left. + 3 \delta_2 X^A \nabla^{[a} (N_{DEF}
      \delta_1 X^D) \nabla^b X^E \nabla^{c]} X^F - \switch \right],
\end{multline}
where the antisymmetrization in the second term is over the tensor
indices only (not the fluid-space indices).  Writing out $\omega^t$ in
terms of our perturbational variables, we have
\begin{multline}
  \label{fRomegat}
  \omega^t = \varrho' \nu e^{2 \Lambda - 2 \Phi} \left( \fpder{\xi_1}{t}
  \xi_2 - \fpder{\xi_2}{t} \xi_1 \right)  \\ + e^{- 2 \Phi} \left( b_1
  \fpder{\lambda_2}{t} - \fpder{b_1}{t} \lambda_2 - b_2
  \fpder{\lambda_1}{t} + \fpder{b_2}{t} \lambda_1 \right)
\end{multline}
where we have defined $b = \delta( f'(\alpha)) = f''(\alpha) \delta
\alpha$.  (Note that the $t$-component of the symplectic current about
a spherically symmetric static solution vanishes in pure Einstein
gravity, i.e., $\omega^t_\text{Ein} = 0$.)

The first main step towards obtaining a variational principle is to
solve the linearized constraints.  To do this, we must calculate the
constraint tensor $C_{ab}$, as defined in \eqref{Cdef}.  Since all the
fields in our theory other than the metric are scalars with respect to
the spacetime metric, the resulting expression is particularly simple:
\begin{equation}
  \label{fRC}
  C_{ab} = 2 (\mathcal{E}_G)_{ab},
\end{equation}
where $(\mathcal{E}_G)_{ab}$, the metric equation of motion, is given
by \eqref{fReineq}.   We can then find algebraic equations for
$\lambda$ and $\pder{\phi}{r}$.   The algebraic equation for $\lambda$
will be given by the solution of  the equation $F = 0$, where $F$ is
given by \eqref{Fdef1} and \eqref{Fdef2}.  Similarly, an algebraic
equation for $\pder{\phi}{r}$ can be found by solving the equation
\begin{equation}
  \delta C_{rr} = (\delta \mathcal{E}_G)_{rr} = 0.
\end{equation}

The situation is thus very similar to the example of pure Einstein
gravity coupled to perfect fluid matter described in \cite{GVP}; only
the precise form of the perturbational equations of motion $F = 0$ and
$(\delta \mathcal{E}_G)_{rr} = 0$ are different.  In the $f(R)$
gravity case, these equations are
\begin{equation}
  \label{fRF}
  F = \mathcal{S} \lambda - \fpder{b}{r} + \fpder{\Phi}{r} b - 8 \pi
  e^{2 \Lambda} \varrho' \nu \xi
\end{equation}
and
\begin{multline}
  \label{fRGrr1}
  (\delta \mathcal{E}_G)_{rr} = e^{- 2 \Lambda}  \mathcal{S}
  \fpder{\phi}{r}  - e^{-2 \Phi} \frac{\partial^2 b}{\partial t^2} -
  e^{-2 \Lambda} \left( \fpder{\Phi}{r} + \frac{2}{r} \right)
  \fpder{b}{r} \\ - e^{-2 \Lambda} \left( 2 \left( \fpder{\Phi}{r} +
  \frac{2}{r} \right)  \mathcal{S} - \frac{6}{r^2} f'(\alpha) \right)
  \lambda \\ + \left( e^{-2 \Lambda} \frac{2}{r} \fpder{\Phi}{r} -
  \frac{1}{r^2} \left(   1 - e^{-2 \Lambda} \right) + \frac{\alpha}{2}
  \right) b \\ - 8 \pi \varrho'' \nu^2 \left( \fpder{\xi}{r} + \left(
  \fpder{\Lambda}{r} +  \frac{2}{r} + \frac{1}{\nu} \fpder{\nu}{r}
  \right) \xi - \lambda \right),
\end{multline}
where we have introduced the new quantity
\begin{equation}
  \label{Sdef}
  \mathcal{S} = \fpder{}{r} f'(\alpha) +\frac{2}{r} f'(\alpha)
\end{equation}
for notational convenience.  Equation \eqref{fRF} then implies that
our algebraic equation for $\lambda$ is
\begin{equation}
  \label{fRlambda}
  \lambda = \mathcal{S}^{-1} \left( \fpder{b}{r} - \fpder{\Phi}{r} b +
  8 \pi e^{2 \Lambda} \varrho' \nu \xi \right).
\end{equation}

We could plug this result into \eqref{fRGrr1} to obtain an algebraic
equation for $\pder{\phi}{r}$;  however, the resulting expression is
somewhat complicated.  In fact, we can derive a simpler expression for
$\pder{\phi}{r}$.  To do so, we note that $(\mathcal{E}_G)_{ab} -
\frac{1}{2} g_{ab} (\mathcal{E}_G)_c {}^c = 0$ if and only if
$(\mathcal{E}_G)_{ab} = 0$.  In the case of $f(R)$ gravity, this
equation is given by
\begin{multline}
  f'(\alpha) R_{ab} - \frac{1}{2} g_{ab} \Box f'(\alpha) - \nabla_a
  \nabla_b f'(\alpha) \\ + \frac{1}{2} g_{ab} \left(f(\alpha) -
  f'(\alpha) \alpha \right) - 8 \pi \left( T_{ab} - \frac{1}{2} g_{ab}
  T^c {}_c \right) = 0.
\end{multline}
Using the trace of this equation to eliminate the $\Box f'(\alpha)$
term, along with the equation $R = \alpha$, we can show that
\begin{multline}
  \label{fRneweq}
  f'(\alpha) R_{ab} + \frac{1}{6} g_{ab} \left(f(\alpha) - 2 f'(\alpha)
  \alpha \right) - \nabla_a \nabla_b f'(\alpha) \\ - 8 \pi \left( T_{ab}
  - \frac{1}{3} g_{ab} T^c {}_c \right) = 0.
\end{multline}
To zero-order, the $\theta \theta$-component of this equation is
\begin{multline}
  \label{fRneweq0}
  \frac{1}{r} f'(\alpha) \left( \fpder{\Lambda}{r} - \fpder{\Phi}{r} +
  \frac{1}{r} \left( e^{2 \Lambda} - 1 \right) \right) \\ + \frac{1}{6}
  e^{2 \Lambda} \left( f(\alpha) - 2 \alpha f'(\alpha) \right) -
  \frac{1}{r} \fpder{}{r}f'(\alpha) - \frac{8 \pi}{3} e^{2 \Lambda}
  \varrho = 0,
\end{multline}
and to first order, it is
\begin{widetext}
\begin{multline}
  \label{fRphipr}
  \frac{1}{r} f'(\alpha) \left( \fpder{\lambda}{r} - \fpder{\phi}{r} +
  \frac{2}{r} e^{2 \Lambda} \lambda \right) + \left( \frac{1}{r} \left(
  \fpder{\Lambda}{r} -   \fpder{\Phi}{r} \right) + \frac{1}{r^2} \left(
  e^{2 \Lambda} - 1 \right) - \frac{1}{6} e^{2 \Lambda} \left(
  \frac{f'(\alpha)}{f''(\alpha)} + 2 \alpha \right)  \right) b -
  \frac{1}{r} \fpder{b}{r} \\ + \frac{1}{3} e^{2 \Lambda} \left(
  f(\alpha) - 2 \alpha f'(\alpha) - 16 \pi \varrho \right) \lambda   -
  \frac{8 \pi}{3} \varrho' \nu e^{2 \Lambda} \left( \fpder{\xi}{r} +
  \left( \fpder{\Lambda}{r} + \frac{2}{r} + \frac{1}{\nu} \fpder{\nu}{r}
  \right) \xi - \lambda \right) = 0.
\end{multline}
\end{widetext}
This equation, with $\lambda$ given by \eqref{fRlambda}, can then be
solved algebraically for $\pder{\phi}{r}$ in terms of the matter
variables and their derivatives.

Our next step is to obtain the reduced matter equations of motion,
i.e., eliminate the metric degrees of freedom $\lambda$ and
$\pder{\phi}{r}$ from the remaining first-order equations of motion.
These remaining equations of motion are the ``scalar'' equation $R -
\alpha = 0$, which at first order is
\begin{multline}
  -\frac{\partial^2 \phi}{\partial r^2} + \fpder{\Phi}{r}
  \fpder{\lambda}{r} + \fpder{\Lambda}{r} \fpder{\phi}{r} - 2
  \fpder{\Phi}{r} \fpder{\phi}{r} + \frac{2}{r} \left(
  \fpder{\lambda}{r} - \fpder{\phi}{r} \right) \\  + \frac{2}{r^{2}}
  e^{2 \Lambda} \lambda + e^{2 \Lambda - 2 \Phi} \frac{\partial^2
    \lambda}{\partial t^2} - e^{2 \Lambda} \left( \alpha \lambda +
  \frac{b}{2 f''(\alpha)} \right) = 0
\end{multline}
and the matter equation of motion, which as in the case of pure
Einstein gravity is
\begin{multline}
  \varrho' \left[ -e^{2 \Lambda - 2 \Phi} \frac{\partial^2 \xi}{\partial
      t^2} + \frac{\partial \phi}{\partial r}  \right] \\ + \left(
  \frac{\partial}{\partial r} + \frac{\partial \Phi}{\partial r}
  \right)   \left[ \varrho'' \nu \left( \frac{\partial \xi}{\partial r}
    + \left( \frac{1}{\nu} \frac{\partial \nu}{\partial r} +
    \frac{\partial \Lambda}{\partial r} + \frac{2}{r} \right) \xi -
    \lambda \right)\right] \\ = 0.
\end{multline}
We can then eliminate the gravitational equations of motion using
\eqref{fRlambda} and \eqref{fRphipr}, leaving equations solely in
terms of the matter variables $b$ and $\xi$.  Equivalently, we can
write our ``reduced'' equations in terms of $b$ and a new variable
$\zeta$, defined by
\begin{equation}
  \label{zetadef}
  \zeta \equiv \xi - \mathcal{S}^{-1} b.
\end{equation}
(The utility of this new variable will become evident when we obtain
the reduced symplectic form.) Performing these algebraic
manipulations, the resulting reduced evolution equation for $\zeta$
takes the form
\begin{equation}
  \label{fRzetaeq}
  e^{2 \Lambda - 2 \Phi} \varrho' \nu \frac{\partial^2 \zeta}{\partial
    t^2} = \mathcal{A}_1 \frac{\partial^2 \zeta}{\partial r^2} +
  \mathcal{A}_2 \fpder{\zeta}{r} +  \mathcal{A}_3 \zeta + \mathcal{A}_4
  \fpder{b}{r} + \mathcal{A}_5 b
\end{equation}
where the coefficients $\mathcal{A}_i$ are dependent on the background
fields.  Similarly, the evolution equation for $b$ takes the form
\begin{equation}
  \label{fRbeq}
  e^{ - 2 \Phi} \frac{\partial^2 b}{\partial t^2} = \mathcal{B}_1
  \frac{\partial^2 b}{\partial r^2} + \mathcal{B}_2 \fpder{b}{r} +
  \mathcal{B}_3 b + \mathcal{B}_4 \fpder{\zeta}{r} + \mathcal{B}_5 \zeta.
\end{equation}
We see from the form of the above equations that $f(R)$ gravity has
cleared another hurdle required to have a valid variational principle:
the reduced equations do in fact take the form \eqref{simpleform},
containing only second derivatives in time and only up to second
radial derivatives.

We can also apply these constraint equations to the symplectic form.
For the symplectic form, we only need to eliminate $\lambda$ using the
first-order constraint equation \eqref{fRlambda}.  Performing an
integration by parts to eliminate the mixed derivatives that result,
and applying the background equations of motion, we obtain our reduced
symplectic form:
\begin{multline}
  \label{fRsympform}
  \Omega = 4 \pi \int \dif r \, r^2 e^{\Lambda + \Phi} \left[ e^{-2
      \Phi} \mathcal{S}^{-2} \frac{6}{r^2} f'(\alpha) \fpder{b_1}{t} b_2
    \right. \\ \left. + e^{2 \Lambda - 2 \Phi} \varrho' \nu
    \fpder{\zeta_1}{t} \zeta_2 \right] - \switch.
\end{multline}
As noted above \eqref{Wdef}, this reduced symplectic form defines a
three-form $\form{W}_{\alpha \beta}$.  For a valid variational
principle to exist, this $\form{W}_{\alpha \beta}$ must be positive
definite in the sense of \eqref{Wposdef}. We can see that the
$\form{W}_{\alpha \beta}$ defined by \eqref{fRsympform} is positive
definite if and only if $f'(\alpha) > 0$ and $\varrho' \nu > 0$ in our
background solutions.  The latter condition will hold for any matter
satisfying the null energy condition, since $\varrho' \nu = \rho + P$;
however, the former condition must be checked in order to determine
whether a valid variational principle exists.\footnote{Should
$f'(\alpha)$ fail to be positive in the background solutions, all is
not necessarily lost;  we can still attempt to analyze the reduced
equations that we have obtained.  See Section \ref{TeVeSsec} for an
example of such an analysis.}  For the particular $f(R)$ chosen by
Carroll \textsl{et al.}  \cite{CDTT}, we have $f'(R) = 1 + \mu^4 / R^2
> 0$, and so $\form{W}_{\alpha \beta}$ is always positive definite in
the required sense for this choice of $f(R)$.

All that remains is to actually write down the variational principle
for $f(R)$ gravity.  As noted above \eqref{basicvarprin}, the
variational principle will take the form
\begin{equation}
  \label{fRsimpform}
  \omega_0^2 \leq \frac{ (\psi, \mathcal{T} \psi)}{(\psi, \psi)}
\end{equation}
where $\mathcal{T}$ is the time-evolution operator.  In our case,
$\psi$ denotes a two-component vector, whose components are the
functions $\zeta$ and $b$.  The inner product in which $\mathcal{T}$
is self-adjoint can be read off from \eqref{fRsympform}, and thus the
denominator of \eqref{fRsimpform} will be
\begin{equation}
  \label{fRdenom}
  (\psi, \psi) = 4 \pi \int \dif r \, r^2 e^{\Lambda - \Phi} \left[
    \mathcal{S}^{-2} f'(\alpha) \frac{6}{r^2} b^2 + e^{2 \Lambda} \varrho'
    \nu \zeta^2 \right].
\end{equation}
The numerator of \eqref{fRsimpform} is, of course, rather more
complicated.  After multiple integrations by parts and applications of
the background equations of motion, we can put this quantity in the
form 
\begin{multline}
  \label{fRnumer}
  (\psi, \mathcal{T} \psi) = 4 \pi \int \dif r \, \left[ \mathcal{C}_1
    \left( \fpder{b}{r} \right)^2  + \mathcal{C}_2 \left( \fpder{\zeta}{r}
    \right)^2 \right. \\ \left. + \mathcal{C}_3 \left( \zeta \fpder{b}{r}
    - b \fpder{\zeta}{r} \right)  + \mathcal{C}_4 b^2  + \mathcal{C}_5
    \zeta^2 + 2 \mathcal{C}_6 b \zeta \right]
\end{multline}
where the $\mathcal{C}_i$ coefficients are functions of the background
fields.  The exact forms of the $\mathcal{C}_i$ are given in Appendix
\ref{fRapp}. 

\subsection{Discussion}

All of our results thus far have been independent of the choice of
$f(R)$ (assuming, of course, that $f''(R) \neq 0$.)  In the case of
Carroll \textsl{et al.}'s $f(R)$ gravity, we can now use this
variational principle to show that this theory is highly unstable for
Newtonian solutions.  For the choice $f(R) = R - \mu^4 / R$, we have
$f'(R) = 1 + \mu^4 / R^2$ and $f''(R) = - 2 \mu^4 / R^3$.  Moreover,
for a quasi-Newtonian stellar interior, we will have $\alpha = R
\approx \rho$, where $\rho$ is the matter density in the star.  In
particular, this implies that
\begin{equation}
  \label{fRapproxvar}
  \mathcal{C}_4 \approx 2 e^{\Phi + \Lambda} \mathcal{S}^{-2}
  \frac{(f'(\alpha))^2}{f''(\alpha)} \approx - e^{\Phi + \Lambda}
  \mathcal{S}^{-2} \frac{\rho^3}{\mu^4}
\end{equation}
since the above term will dominate over all the others in
$\mathcal{C}_4$.  (Note that for a star with the density of the Sun,
and the choice of $\mu \approx 10^{-27} \text{ m}^{-1}$ made by
Carroll \textsl{et al.}, $\rho/\mu^2 \approx 10^{10}$.)  This implies
that for a test function $\psi$ with $\zeta$ set to zero, we will have
\begin{equation}
  \omega_0^2 \lesssim   \frac{\int \dif r \, e^{\Lambda + \Phi}
    \mathcal{S}^{-2} \left[ - \frac{\rho^3}{\mu^4} b^2 + 6 e^{-2 \Lambda}
      \left( \fpder{b}{r} \right)^2 \right] }{ 6 \int \dif r \, e^{\Lambda -
      \Phi} \mathcal{S}^{-2} b^2 }
\end{equation}
(note that $f'(\alpha) = 1 + \mu^4/\rho^2 \approx 1$ in the stellar
interior.)   As a representative mass distribution, we take the
Newtonian mass profile of an $n=1$ polytrope:
\begin{equation}
  \rho(r) = \rho_0 \frac{ R \sin \left( \frac{r}{R} \right)}{r}
\end{equation} 
where $R$ is the radius of the star and $\rho_0$ is its central
density.  We take $\rho_0$ to be of a typical stellar density, $\rho_0
\approx 10^{-24} \text{ metres}^{-2}$.  Numerically integrating
\eqref{fRapproxvar} with a test function of the form
\begin{equation}
  \label{fRtestfn}
  b = \begin{cases} 1 - \frac{r}{R} & r \leq R \\ 0 & r > R \end{cases},
\end{equation}
we find that
\begin{equation}
  \label{fRbound}
  \omega_0^2 \lesssim -9 \times 10^{35} \text{ metres}^{-2}
\end{equation}
which corresponds to an instability timescale of $\tau \approx 4
\times 10^{-27}$ seconds.  This timescale is of the same magnitude as
the instability found in \cite{DolKaw}.

We see, then, that for the $f(R)$ originally chosen by Carroll
\textsl{et al.}, the theory is extremely unstable in the presence of
matter.  Moreover, a similar argument will obtain for any choice of
$f(R)$ for which quasi-Newtonian solutions exist and for which
$f''(R)$ is sufficiently small and negative at stellar-density
curvature scales.  We can always pick a test function $b(r)$ lying
entirely inside the stellar interior such that $\pder{b}{r}$ is of
order $b/R$.  Thus, if $\left| f'(\rho) / f''(\rho) \right| \gg 1/R^2$
for a typical stellar density $\rho$, the $b^2$ term in
\eqref{fRapproxvar} will dominate the $(\pder{b}{r})^2$ term,  and the
resulting lower bound on $\omega_0^2$ will then be of order
$f'(\alpha) / f'' (\alpha)$.  If the choice of $f(\alpha)$ results in
this quantity being negative, quasi-Newtonian stellar solutions will
be unstable in the corresponding theory.  We can thus rule out any
theory (e.g., \cite{CDTT, Zhang}) for which $f''(\rho)$ is
sufficiently small and negative.

This result seems to be closely related, if not identical to, the
``high-curvature'' instabilities found in cosmological solutions by
Song, Hu, and Sawicki \cite{SoHuSa}. Particularly telling is the fact
that the instability time scale found in their work is proportional to
$\sqrt{|f''(R)/f'(R)|}$, the same time-scale found in the present
work.

\section{Einstein-{\ae}ther theory \label{aethersec}}

\subsection{Theory}

Einstein-{\ae}ther theory \cite{Aether1,Aetherwave} was first
formulated as a toy model of a gravitational theory in which Lorentz
symmetry is dynamically broken.  This theory contains, along with the
metric $g_{ab}$, a vector field $u^a$ which is constrained (via a
Lagrange multiplier $Q$) to be unit and timelike.  The Lagrangian
four-form for this theory is
\begin{multline}
  \label{EAlag}
  \form{\mathcal{L}} = \left( \frac{1}{16 \pi} R + K^{ab} {}_{cd}
  \nabla_a u^c \nabla_b u^d + Q (u^a u_a + 1) \right) \form{\epsilon} \\
  + \form{\mathcal{L}}_\text{mat} \left[ A, g^{ab}, u^a \right]
\end{multline}
where $A$ denotes any matter fields present in the theory, and
\begin{equation}
  \label{Kdef}
  K^{ab} {}_{cd} = c_1 g^{ab} g_{cd} + c_2 \delta^a {}_c \delta^b {}_d +
  c_3 \delta^a {}_d \delta^b {}_c - c_4 u^a u^b g_{cd}.
\end{equation}
The $c_i$ constants determine the strength of the vector field's
coupling to gravity, as well as its dynamics.\footnote{Note that our
definitions of the coefficients $c_i$ differ from those in
\cite{Aether1,Aetherwave,Aethersph,AetherBH}, as do the respective
metric signature conventions.  The net result is that to translate
between our results and those of the above papers, one must flip the
signs of all four coefficients.}  (Note that in the case $c_3 = - c_1
> 0$ and $c_2 = c_4 = 0$, we have the conventional kinetic term for a
Maxwell field;  this special case was examined in \cite{Kost1}, prior
to Jacobson and Mattingly's work.) In the present work, we will work
in the ``vacuum theory'', i.e., in the absence of matter fields $A$.

Performing the variation of the Lagrangian four-form, we find that
\begin{equation}
  \label{EAvar}
  \delta \form{\mathcal{L}} = \left( (\mathcal{E}_G)_{ab} \delta
  g^{ab} + (\mathcal{E}_u)_a \delta u^a + \mathcal{E}_Q \delta Q +
  \nabla_a \theta^a \right) 
  \form{\epsilon} 
\end{equation}
where
\begin{subequations}
\begin{multline}
  \label{EAeineq}
  (\mathcal{E}_G)_{ab} = \frac{1}{16 \pi} G_{ab} + \nabla_c \left(
  J^c {}_{(a} u_{b)} + J_{(ab)} u^c - J_{(a} {}^c u_{b)} \right) \\ + c_1
  \left( \nabla_a 
  u^c \nabla_b u_c - \nabla^c 
  u_a \nabla_c u_b \right) + c_4 \dot{u}_a \dot{u}_b \\ - Q u_a u_b -
  \frac{1}{2} g_{ab} J^c {}_d \nabla_c u^d,
\end{multline}
\begin{equation}
  \label{EAveceq}
  (\mathcal{E}_u)_a = -2 \nabla_b J^b {}_a - 2 c_4 \dot{u}_b \nabla_a
  u^b + 2 Q u_a, 
\end{equation}
\begin{equation}
  \label{EAQeq}
  (\mathcal{E}_Q) = u^a u_a + 1,
\end{equation}
\end{subequations}
and
\begin{equation}
  \label{EAtheta}
  \theta^a = \theta^a_\text{Ein} + 2 J^a {}_b \delta u^b + \left( J_b {}^a u_c
  - J^a {}_b u_c - J_{bc} u^a \right) \delta g^{bc}.
\end{equation}
In the above, we have introduced the notation $\dot{u}^a = u^b
\nabla_b u^a$ and $J^a {}_b = K^{ac} {}_{bd} \nabla_c u^d$;
$\theta^a_\text{Ein}$ is defined by \eqref{thetaein}, as above.
Equation \eqref{EAQeq} is the constraint that $u^a$ is unit and 
timelike, while \eqref{EAeineq} and \eqref{EAveceq} are the equations
of motion for $g^{ab}$ and $u^a$, respectively.  If desired, we can
eliminate the Lagrange multiplier $Q$ from these equations by
contracting \eqref{EAveceq} with $u^a$, resulting in the equation
\begin{equation}
  \label{Qeq}
  Q = - u^a \nabla_b J^b {}_a - c_4 u^a \dot{u}_b \nabla_a u^b.
\end{equation}

We now take the variation of the symplectic potential current
to obtain the symplectic current.  The resulting expression can be
written in the form 
\begin{equation}
  \label{EAomega}
  \omega^a = \omega^a_\text{Ein} + \omega^a_\text{vec}
\end{equation}
where $\omega^a_\text{Ein}$ is the usual symplectic current for
pure Einstein gravity (given by \eqref{omegaein}), and
$\omega^a_\text{vec}$ is obtained by taking the antisymmetrized
variation of the last two terms in \eqref{EAtheta}.  The exact form
of this tensor expression is rather complicated, and can be found in
Appendix \ref{EAapp}.

\subsection{Obtaining a variational principle}

As before, we are primarily concerned with the $t$-component of
$\omega^a$, and its form in terms of the perturbational variables.  We
will use the usual spherical gauge \eqref{spheremet} for
$g^{ab}$.  We will further assume that $u^a \propto t^a$ in the
background solution (the so-called ``static {\ae}ther'' assumption),
and that to first order we have 
\begin{equation}
\label{EAupsdef}
u^\mu = (e^{-\Phi} - \phi e^{-\Phi}, \upsilon, 0, 0),
\end{equation}
i.e., the first-order perturbation to $u^t$ is $- \phi e^{-\Phi}$
while the first-order perturbation to $u^r$ is $\upsilon$.  (Note that
under these conventions, $g_{ab} u^a u^b = -1$ to first order, as
required.) The background equations of motion can then be calculated
to be 
\begin{multline}
  \label{EAGtt0}
  (\mathcal{E}_G)_{tt} = e^{2\Phi - 2 \Lambda} \left[ \frac{2}{r}
  \fpder{\Lambda}{r} + 
  \frac{1}{r^2} \left( e^{2 \Lambda} - 1 \right) \right. \\ \left. +
  c_{14} \left( 
  \fpdert{\Phi}{r} + \frac{1}{2} \left( \fpder{\Phi}{r} \right)^2 -
  \fpder{\Phi}{r} \left( \fpder{\Lambda}{r} - \frac{2}{r} \right)
  \right) \right]
\end{multline}
and
\begin{equation}
  \label{EAGrr0}
  (\mathcal{E}_G)_{rr} = \frac{2}{r} \fpder{\Phi}{r} - \frac{1}{r^2}
  \left( e^{2 \Lambda} - 1 \right) - \frac{c_{14}}{2} \left(
  \fpder{\Phi}{r} \right)^2,
\end{equation}
where we have defined $c_{14} = c_1 + c_4$ and used the equation
\begin{multline}
  \label{EAQ0}
  Q = e^{-2 \Lambda} \left[ c_3 \fpder{\Phi}{r} \left(
  \fpder{\Lambda}{r} - \frac{2}{r} \right) \right. \\ \left. - (c_1 +
  c_3 + 2 c_4) \left( 
  \fpder{\Phi}{r} \right)^2 - c_3 \fpdert{\Phi}{r} \right]
\end{multline}
to eliminate $Q$.  The final equation of motion, $(\mathcal{E}_u)_a =
0$, is satisfied trivially if the {\ae}ther is static.

In terms of these variables, the $t$-component of $\omega^a$ can be
calculated to be
\begin{multline}
  \label{EAomegat}
  \omega^t = 2 e^{-2\Phi} \left[ c_{123} \fpder{\lambda_1}{t} \lambda_2 +
    c_{123} e^{\Phi} \fpder{\upsilon_1}{r} \lambda_2
    \right. \\ + e^{\Phi} \left( c_{123} \fpder{\Lambda}{r} + (c_{123}- c_{14})
    \fpder{\Phi}{r} + c_2 \frac{2}{r} \right) \upsilon_1 \lambda_2 \\
    \left. + c_{14} e^{\Phi} \upsilon_1 \fpder{\phi_2}{r} + e^{2
      \Lambda} c_{14} \upsilon_1 \fpder{\upsilon_2}{t} \right] - \switch
\end{multline}
where we have defined $c_{123} = c_1 + c_2 + c_3$.  In what follows,
we will assume that $c_{14} \neq 0$, and, except where otherwise
noted, that $c_{123} \neq 0$ as well.

Our next step, as usual, is to solve the constraints.  However, in
this theory we have the added complication of the presence of a
vector field.  This means, in particular, that the tensor $C_{ab}$ (as
defined in \eqref{Cdef}) is not merely proportional to
$(\mathcal{E}_G)_{ab}$, as in the previous section, but is instead
\begin{align}
C_{ab} &= 2 (\mathcal{E}_G)_{ab} + u_a (\mathcal{E}_u)_b \\
&= \nonumber \frac{1}{8 \pi} G_{ab} + 2 \nabla_c \left(
  J_{(ab)} u^c - J_{(a} {}^c u_{b)} \right) \\ \nonumber & \quad + 2
  \left(\nabla_c J^c {}_{[a} 
    u_{b]} + \nabla_c u_{(a} J^c {}_{b)} \right) \\ \nonumber & \quad + 2 c_1
  \left( \nabla_a u^c \nabla_b u_c - \nabla^c 
  u_a \nabla_c u_b \right) \\ & \quad - 2 c_4 \left(u_a \dot{u}_c
  \nabla_b u^c - 
  \dot{u}_a \dot{u}_b \right) - g_{ab} J^c {}_d \nabla_c u^d.
\end{align}
Writing out $\delta C_{rt}$ in terms of our perturbational
quantities, we find that it is indeed a total time derivative, with
\begin{multline}
  \label{EAFeq}
  F = 2 r^2 e^{\Phi - \Lambda} \left( \left( \frac{2}{r} - c_{14}
  \fpder{\Phi}{r} \right) \lambda \right. \\ \left.  + c_{14} e^{2
    \Lambda -   \Phi} \frac{\partial 
    \upsilon}{\partial t} + c_{14} \frac{\partial \phi}{\partial r}
  \right).
\end{multline}
Note that this quantity depends on $\phi$; $\phi$ and its derivatives
can appear in the preconstraint equation $F=0$ when our background
tensor fields have non-vanishing $t$-components, as noted in Footnote
\ref{phifoot}. 

The remaining non-trivial equations of motion are $(\delta
\mathcal{E}_G)_{rr} = 0$ and $(\delta \mathcal{E}_u)_r = 0$;  in terms
of the perturbational variables, these are
\begin{multline}
\label{EAGrreq1}
(\delta \mathcal{E}_G )_{rr} =  - \frac{2}{r^2} e^{2\Lambda} \lambda +
  c_{123} e^{2 \Lambda - 2\Phi} \frac{\partial^2 
  \lambda}{\partial t^2} \\ + \left( \frac{2}{r} - c_{14} \fpder{\Phi}{r}
  \right) \fpder{\phi}{r} + c_{123} e^{2\Lambda - \Phi}
  \frac{\partial^2 \upsilon}{\partial r \partial t} \\
  + e^{2 \Lambda - \Phi} \left( \frac{2 c_2}{r} + c_{123}
  \fpder{\Lambda}{r} + (c_{123} - c_{14})  \fpder{\Phi}{r} \right)
  \fpder{\upsilon}{t}
\end{multline}
and
\begin{widetext}
\begin{multline}
\label{EAveceq1}
\frac{1}{2}(\delta \mathcal{E}_u)_r = c_{14} e^{2 \Lambda -
  2 \Phi} \frac{\partial^2 \upsilon}{\partial t^2} - c_{123} \frac{\partial^2
  \upsilon}{\partial r^2} - c_{123} e^{-\Phi} \frac{\partial^2
  \lambda}{\partial r \partial t} + c_{14} e^{-\Phi} \frac{\partial^2
  \phi}{\partial r \partial 
  t} \\ - c_{123} \left( \frac{2}{r} + \fpder{\Lambda}{r} +
  \fpder{\Phi}{r} \right) \fpder{\upsilon}{r} + e^{-\Phi} \left(
  (c_{123} - c_{14}) \fpder{\Phi}{r} - (c_1 + c_3) \frac{2}{r} \right)
  \fpder{\lambda}{t} \\ - \left[ (c_{123} - c_{14} ) \frac{\partial^2
  \Phi}{\partial r^2} + c_{123} \frac{\partial^2 \Lambda}{\partial
  r^2} + c_{14} \fpder{\Lambda}{r} \fpder{\Phi}{r} + \frac{2}{r}
  \left( (c_1 + c_3) \fpder{\Lambda}{r} + (c_3 - c_4) \fpder{\Phi}{r}
  \right) - c_{123} \frac{2}{r^2} \right] \upsilon.
\end{multline}
\end{widetext}
While we could follow the methods outlined in Section \ref{GVPsec}
to reduce these equations to the basic form
\eqref{simpleform}, it is actually simpler to pursue a different path.
If we solve \eqref{EAFeq} for $\pder{\phi}{r}$, rather than $\lambda$
as usual, and plug the resulting expressions into \eqref{EAGrreq1} and
\eqref{EAveceq1}, there result the equations
\begin{equation}
  \label{modEAGrreq1}
  \fpder{\psi}{t} - e^{2 \Phi - 2 \Lambda} \frac{2}{r^2} \left(
  \frac{2}{c_{14}} + 1 \right) \lambda = 0
\end{equation}
and
\begin{multline} 
  \label{modEAveceq1}
  \fpder{\psi}{r} + \left( \frac{c_1 + c_3 + 1}{c_{123}} \frac{2}{r} -
  \fpder{\Phi}{r} \right) \psi \\ - \frac{(c_1 + c_3 + 1)(c_{123} +
    2 c_2 - 2)}{c_{123}} \upsilon = 0,
\end{multline}
where we have introduced the new variable $\psi$, defined as
\begin{multline}
  \label{EApsidef}
  \psi = c_{123} \left( \fpder{\lambda}{t} + e^{\Phi}
  \fpder{\upsilon}{r} \right) \\ + \left(c_{123} \left(
  \fpder{\Lambda}{r} + \fpder{\Phi}{r} \right) + (c_2 - 1)
  \frac{2}{r^2} \right) e^\Phi \upsilon.
\end{multline}
Equations \eqref{modEAveceq1} and \eqref{EApsidef} can then be
combined to eliminate any explicit $\upsilon$ terms:
\begin{multline}
  \fpder{\lambda}{t} + \frac{r^2}{a} \left[ \frac{c_{123}}{2} \left(
  \fpdert{\psi}{r} + \left( \fpder{\Lambda}{r} - 
  \fpder{\Phi}{r} + \frac{4}{r} \right) \fpder{\psi}{r}
  \right)\right. \\ \left. 
  + \left( - c_{123} \fpder{\Lambda}{r} \fpder{\Phi}{r} +
  \left(\frac{c_{123}}{c_{14}} + c_{123} - c_2 + 1 \right) \frac{1}{r}
  \fpder{\Lambda}{r} \right. \right. \\ \left. \left. + \left(
  \frac{c_{123}}{c_{14}} - c_2 + 1 \right) 
  \frac{1}{r} \fpder{\Phi}{r} \right) \psi \right] = 0 
\end{multline}
where we have defined the constant $a$ to be
\begin{equation}
  \label{EAadef}
  a = (c_1 + c_3 + 1)(c_{123} + 2 c_2 - 2).
\end{equation}
We can then use this equation and \eqref{modEAGrreq1} to write down a
single second-order time-evolution equation for $\psi$:
\begin{multline}
  \label{EAredeq}
  e^{2\Lambda - 2 \Phi} r^2 \frac{c_{14}}{2 + c_{14}}
  \fpdert{\psi}{t} \\ + \frac{r^2}{a} \left[ c_{123} \left(
  \fpdert{\psi}{r} + \left( \fpder{\Lambda}{r} - 
  \fpder{\Phi}{r} + \frac{4}{r} \right) \fpder{\psi}{r} \right)
  \right. \\ \left.
  + \left( - c_{123} \fpder{\Lambda}{r} \fpder{\Phi}{r} +
  \left(\frac{c_{123}}{c_{14}} + c_1 + c_3 + 1 \right) \frac{1}{r}
  \fpder{\Lambda}{r} \right. \right. \\ \left. \left. + \left(
  \frac{c_{123}}{c_{14}} - c_2 + 1 \right) 
  \frac{1}{r} \fpder{\Phi}{r} \right) \psi \right] = 0.
\end{multline}
Moreover, using \eqref{EAFeq} to eliminate the
$\pder{\phi}{r}$ term from \eqref{EAomega}, we find that
\begin{equation}
  \label{EAredomega}
  \omega^t = 2 c_{123} e^{-2 \Phi} (\lambda_2 \psi_1 - \lambda_1 \psi_2)
\end{equation}
and the reduced symplectic form can be written as
\begin{multline}
  \label{EArefsympform}
  \Omega(\psi_1, \psi_2) \\ = - 4 \pi \frac{c_{123} c_{14}}{2 + c_{14}}
  \int_0^\infty \dif r \, r^4 e^{3 \Lambda   - 3 \Phi} 
  \left( \fpder{\psi_1}{t} \psi_2 - \fpder{\psi_2}{t} \psi_1 \right)
\end{multline}
The form $\form{W}_{\alpha \beta}$ thus defined may be positive or
negative definite, depending on the signs of $c_{123}$ and $c_{14}$
(recall that we are assuming that $c_{123}$ and $c_{14}$ are
non-vanishing);  however, it is never indefinite.  Thus, the
symplectic form is in fact of the required 
form \eqref{Wdef}, and the equations of motion can be put in the form
\eqref{simpleform}.  We can therefore write down our variational
principle of the form \eqref{basicvarprin};  the denominator is
\begin{equation}
  \label{EAdenom}
  (\psi, \psi) = - 4 \pi \frac{c_{123} c_{14}}{2 + c_{14}} \int_0^\infty
  \dif r \, r^4 e^{3 \Lambda - 3 \Phi} \psi^2 
\end{equation}
and the numerator is
\begin{widetext}
\begin{multline}
  \label{EAnumer}
  (\psi, \mathcal{T} \psi) =  - 4 \pi \frac{c_{123}}{ a} \int_0^\infty
  \dif r \, 
  r^4 e^{\Lambda - \Phi} \left[ c_{123} \left( \fpder{\psi}{r}
  \right)^2 \right. \\ \left. +
  \left(c_{123} \fpder{\Lambda}{r} \fpder{\Phi}{r} - \left(
  \frac{c_{123}}{c_{14}} - c_2 + 1 \right) 
  \frac{1}{r} \fpder{\Phi}{r} -
  \left(\frac{c_{123}}{c_{14}} + c_1 + c_3 + 1 \right) \frac{1}{r}
  \fpder{\Lambda}{r}  \right) \psi^2 \right].
\end{multline}
\end{widetext}

\subsection{Discussion}

We can now examine the properties of this variational principle to
determine the stability of Einstein-{\ae}ther theory.  The simplest
case to examine is that of flat space.  Let
us denote the coefficient of $\psi^2$ in \eqref{EAnumer} by $Z(r)$,
i.e., 
\begin{multline}
  \label{EAZdef}
  Z(r) = c_{123} \fpder{\Lambda}{r} \fpder{\Phi}{r}  - \left(
  \frac{c_{123}}{c_{14}} - c_2 + 1 \right) 
  \frac{1}{r} \fpder{\Phi}{r} \\ -
  \left(\frac{c_{123}}{c_{14}} + c_1 + c_3 + 1 \right) \frac{1}{r}
  \fpder{\Lambda}{r}.
\end{multline}
In the case of flat spacetime, $Z(r)$ vanishes, and we are left with
\begin{equation}
  \label{EAvarflat}
  \omega_0^2 \leq \frac{ c_{123} (2 + c_{14}) }{c_{14} a} \frac{
  \int_0^\infty \dif r \, r^4  \left(
  \fpder{\psi}{r} \right)^2}{\int_0^\infty
  \dif r \, r^4 \psi^2}.
\end{equation}
Since both integrands are strictly positive, we conclude that flat
space is stable to spherically symmetric perturbations in
Einstein-{\ae}ther theory if and only if 
\begin{equation}
  \label{EAflatstab}
  \frac{ c_{123} (2 + c_{14}) }{c_{14} (c_1 + c_3 + 1)(c_{123} + 2 c_2
  - 2)} > 0. 
\end{equation}
This is the same condition found in \cite{Aetherwave} for the
stability of the ``trace'' wave mode (modulo the aforementioned sign
differences).  The other stability conditions found in
\cite{Aetherwave} --- those corresponding to the ``transverse
traceless metric'' and ``transverse {\ae}ther'' wave modes --- are
incompatible with spherical symmetry, and thus our analysis cannot
reproduce these conditions.

\begin{figure}
  \includegraphics[width=\figwidth]{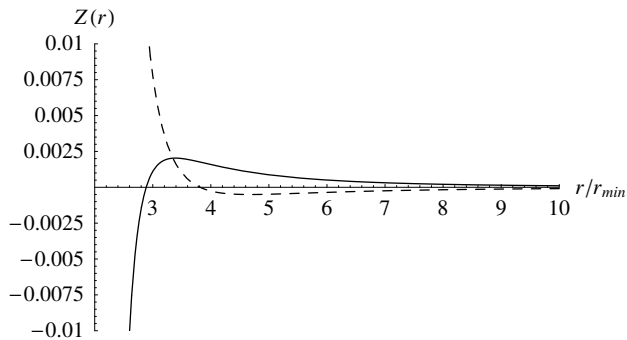}
  \caption{\label{Zfig} Representative plots of $Z(r)$ versus $r$ for
    the solutions described in \cite{Aethersph}.  The solid and dashed
    lines correspond to $c_{123} = \pm 0.1$, respectively;  in both
    cases, $c_2 = 0$ and $c_{14} = - 0.1$.  Note that the choice of
    parameters leading to the dashed plot would lead to instability in
    the asymptotic region.}
\end{figure}

More broadly, we can also analyze the stability of the general
spherically symmetric solutions described in \cite{Aethersph}.  These
(exact) solutions are described in terms of a parameter $Y$:
\begin{equation}
  r(Y) = r_\text{min} \frac{Y - Y_-}{Y} \left( \frac{Y - Y_-}{Y -
    Y_+} \right)^{\frac{1}{2 + Y_+}}
\end{equation}
\begin{equation}
  \Phi(Y) = - \frac{Y_+}{2(2 + Y_+)} \ln \left( \frac{1 - Y/Y_-}{1 -
  Y/Y_+} \right)
\end{equation}
\begin{equation}
  \Lambda(Y) = \frac{1}{2} \ln \left( - \frac{c_{14}}{8} (Y - Y_+) (Y
  - Y_-) \right) 
\end{equation}
where the constants $Y_\pm$ are given by
\begin{equation}
  Y_\pm = - \frac{4}{c_{14}} \left( -1 \pm \sqrt{1 + \frac{c_{14}}{2}}
  \right) 
\end{equation}
and $r_\text{min}$ is a constant of integration related to the mass
$M$ via  
\begin{equation}
  r_\text{min} = \frac{2 M}{- Y_+} (- 1 - Y_+)^{(1 + Y_+)/(2+Y_+)}
\end{equation}
We can then obtain $Z(Y)$ by writing out \eqref{EAZdef} in terms of
these functions of $Y$ (noting that, for example, $\pder{\Lambda}{r} =
(\pder{\Lambda}{Y})/(\pder{r}{Y})$), and then plot $Z$ and $r$
parametrically.  The resulting function is
shown in Figure \ref{Zfig}, for $c_{123} = \pm c_{14}$ and $c_{14} =
-0.1$.  

The first thing we see is that in the asymptotic region, the sign of
$Z(r)$ is determined by the sign of $c_{123}$.  To quantify this, we
can obtain an asymptotic expansion of $Z(r)$ as $r \to \infty$, noting
that $r \to \infty$ as $Y \to 0$.  Doing this, we find that to
leading order in $M/r$, 
\begin{equation}
  \label{EAZasymp}
  Z(r) = c_{123} \frac{M}{r^3} + \dots
\end{equation}
Thus, the coefficient of the $\psi^2$ term in \eqref{EAnumer} is, to
leading order in the asymptotic region, the same sign as that of the
$(\pder{\psi}{r})^2$ 
term.  We can therefore conclude that for a test function in the
``asymptotic region'' of one of these solutions (i.e., $r \gg M$), the
conditions on the $c_i$'s are the same as those in flat
spacetime.\footnote{ Note that if the leading-order coefficient in
  \eqref{EAZasymp} had been different from that of the
  $(\fpder{\psi}{r})^2$ term in \eqref{EAnumer}, we would have
  obtained a new constraint on the $c_i$ coefficients;  the matching
  of these two coefficients seems to be coincidental.}
For a realistic spacetime, of course, there will be some ball of
matter in the central region, and the region where the vacuum solution
holds will be precisely the region where $r \gg M$;  thus, we can
conclude that for a normal star in Einstein-{\ae}ther theory, the
exterior is stable if and only if \eqref{EAflatstab} holds.  
(Henceforth, we will assume that the $c_i$ coefficients have been
chosen with this constraint in mind, unless otherwise specified.)

We also note that as
$r$ approaches $r_\text{min}$, $Z(r)$ diverges negatively.  One might
ask whether a test function in this region (in a spherically symmetric
spacetime surrounding a compact object, say) could lead to an
instability.  We investigated this question numerically using our
variational principle;  however, our results for $(\psi, \mathcal{T}
\psi)$ were positive for all test functions $\psi$ that we
tried.  Roughly speaking, the derivative terms in \eqref{EAdenom}
always won out over the effects of the negative $Z(r)$.\footnote{We
  have been assuming a ``static {\ae}ther'' 
  throughout this work;  however, there also exist solutions in which
  the {\ae}ther vector $u^a$ is not globally aligned with the
  time-translation vector field \cite{AetherBH}.  Thus, even supposing
  that such ``extremely compact stars'' are phenomenologically
  realistic, and that the exterior solutions for such objects are
  unstable, these ``non-static {\ae}ther'' solutions might still be
  physically viable.}
This is, of course, far from a definitive proof of the positivity of
\eqref{EAnumer};  and it should be emphasized that the above analysis
has not considered the effects of matter on the stability of such
solutions.  Nevertheless, the above results are at least indicative
that the spherically symmetric vacuum solutions Einstein-{\ae}ther
theory do not possess any serious stability problems.

\subsection{The case of $c_{123} = 0$}

The above analysis
assumed that $c_{123}$ was non-vanishing;  however, the action
originally considered by Jacobson and Mattingly \cite{Aether1} used
a ``Maxwellian'' action, i.e., the kinetic terms for the {\ae}ther
field were of the form $F_{ab} F^{ab}$, corresponding to $c_3 = -c_1$
and $c_2 = 0$.  This case is therefore of some interest.  

In the preceding analysis, the perturbational equations of motion 
\eqref{EAGrreq1} and \eqref{EAveceq1} were obtained without any
assumptions concerning $c_{123}$, as was the preconstraint equation
\eqref{EAFeq}.  We can therefore simply set $c_{123}$ to zero in these
equations and use \eqref{EAFeq} to solve for $\pder{\phi}{r}$, as we
did in the general case.  After application of the background
equations of motion, there result the equations
\begin{equation}
  \label{EAc0eq1}
  - \frac{2}{r^2} \left( \frac{2}{c_{14}} + 1 \right) \lambda + e^{2
    \Lambda - \Phi} \frac{2}{r} (c_2 - 1) \fpder{\upsilon}{t} = 0
\end{equation}
and
\begin{equation}
  \label{EAc0eq2}
  - e^{-\Phi} \frac{2}{r} (1 - c_2) \fpder{\lambda}{t} - Z_0 \upsilon =
  0,
\end{equation}
where we have defined $Z_0$ in terms of the background fields:
\begin{multline}
  Z_0 =   \frac{c_{14}}{2} \left( \fpder{\Phi}{r} \right)^2 - 
  \frac{2}{r} c_2 \left( \fpder{\Lambda}{r} + \fpder{\Phi}{r}
  \right) \\ + \frac{2}{r} \fpder{\Lambda}{r} +
  \frac{1}{r^2} \left( e^{2 \Lambda} - 1 \right).
\end{multline}
We can then combine these two equations to make a single second-order
equation for $\upsilon$:
\begin{equation}
  \label{EAc0upseq}
  e^{2 \Lambda - 2 \Phi} \frac{c_{14} (c_2 - 1)^2}{2 + c_{14}}
  \fpdert{\upsilon}{t} - Z_0 \upsilon = 0.
\end{equation}
Since there are no radial derivatives to contend with, the solutions to
this equation are all of the form
\begin{equation}
  \label{EAc0form}
  \upsilon(r,t) = f(r) \exp \left[ \pm \sqrt{Z_0(r) \left(
  \frac{2}{c_{14}} + 1 \right) } e^{\Phi - \Lambda}
  |c_2 - 1| t \right]
\end{equation}
where $f(r)$ is an arbitrary function of $r$.

\begin{figure}
  \includegraphics[width=\figwidth]{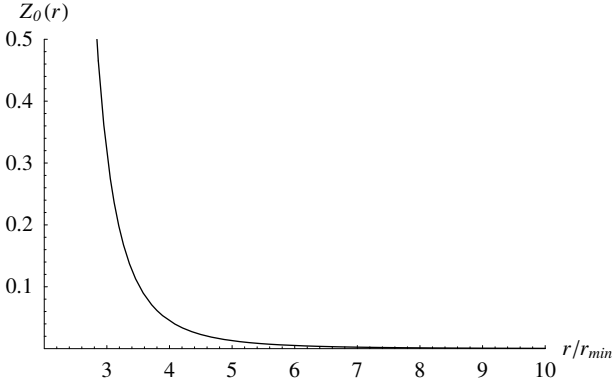}
  \caption{ \label{Z0fig} Plot of $(\frac{2}{c_{14}} + 1 ) Z_0(r)$,
    relevant for Einstein-{\ae}ther theory with $c_{123} = 0$.  In
    this plot, we have chosen $c_{14} = -0.1$ and $c_2 = 0.1$. }
\end{figure} 
Stability of these solutions will thus depend on the sign of the
quantity $(\frac{2}{c_{14}} + 1)Z_0(r)$.  We can
plot $Z_0(r)$ parametrically, as was done in the general case;  the
resulting function, shown in Figure \ref{Z0fig}, falls off rapidly in
the asymptotic region.  We can also perform an asymptotic expansion
similar to that done in the general case to find the large-$r$
behaviour of $Z_0(r)$;  the result is that
\begin{equation}
  Z_0(r) \approx (1-c_2) c_{14} \frac{M^2}{r^4} + \dots
\end{equation}
We conclude that in the $c_{123} = 0$ case, the spherically symmetric
static solutions of Einstein-{\ae}ther theory are unstable unless 
\begin{equation}
  \label{EAc0cond}
  (c_2 - 1) (c_{14} + 2) > 0,
\end{equation}
i.e., unless either $c_2 > 1 $ and $c_{14} > -2$, or $c_2 < 1$ and
$c_{14} < -2$.  Note that this excludes the limit of ``small $c_i$''
often considered in works such as \cite{Aetherwave}.  In the
``Maxwellian'' case, this instability is likely related to the
non-boundedness of the Hamiltonian \cite{Clayton};  however, our
analysis also suggests that non-standard kinetic terms can in fact
stabilize the theory (at least in the timelike {\ae}ther case),
contrary to the arguments made in that work.

Finally, we note that even if \eqref{EAc0cond} holds, the
perturbational solutions of the form \eqref{EAc0form} may still be
problematic, as the radial gradients of $\upsilon$ will grow linearly
with time.  To quantify what magnitude of gradients would be acceptable,
we note that in the $c_{123} = 0$ case, there is only one dynamical
degree of freedom;  thus, $\upsilon$ uniquely determines $\lambda$ and
$\pder{\phi}{r}$.  Setting $c_{123} = 0$ in \eqref{EAFeq} and
\eqref{EAGrreq1} allows us to solve these equations for $\lambda$ and
$\pder{\phi}{r}$;  in particular, 
\begin{multline}
  \fpder{\phi}{r} = -\left[ \frac{2}{r^2} c_{14} e^{2 \Lambda} +
  \left( \frac{2}{r} - c_{14} \fpder{\Phi}{r} \right) \left(
  \frac{2}{r} c_2 - c_{14} \fpder{\Phi}{r} \right) \right] \\
  \times \left[ \frac{2}{r^2} c_{14} e^{2 \Lambda} +
  \left( \frac{2}{r} - c_{14} \fpder{\Phi}{r} \right)^2 \right]^{-1}
  e^{2 \Lambda - \Phi} \fpder{\upsilon}{t},
\end{multline}
which simplifies in a nearly-flat spacetime to be
\begin{equation}
  \fpder{\phi}{r} \approx - \frac{c_{14} + 2 c_2}{c_{14} + 2}
  \fpder{\upsilon}{t}.
\end{equation}
Applying this to our solution \eqref{EAc0upseq}, we find that for
sufficiently large $t$
\begin{equation}
  \left| \fpdert{\phi}{r} \right| \approx 2 \sqrt{(c_2 - 1)^3 (c_{14} + 2
  c_2)} \frac{M}{r^3} \left| \fpder{\phi}{r} \right| t
\end{equation}
We can estimate typical scales for $\pder{\phi}{r}$ by looking at
other sources, such as planets.  The scale of perturbations due to the
Earth's gravitational field (at its surface) is given by $\pder{\phi}{r}
\approx M_\oplus / r_\oplus^2$, where $M_\oplus$ and $r_\oplus$ are
the mass and radius of the Earth, respectively.  Thus, the perturbation
to $\partial^2 \phi / \partial r^2$ near the Earth's surface will be
of the order 
\begin{equation}
  \left| \fpdert{\phi}{r} \right| \approx 2 \sqrt{(c_2 - 1)^3 (c_{14} + 2
  c_2)} \frac{M_\odot M_\oplus}{R^3 r_\oplus^2} t
\end{equation}
where $R$ is the radius of Earth's orbit and $M_\odot$ is the mass of
the Sun.  This enhancement to $\partial^2 \phi / \partial r^2$ will
lead to an observable change in the tidal effects due to the Sun's
gravity;  demanding that these remain small relative to the normal
Newtonian tidal effects, i.e., $\partial^2 \phi / \partial r^2 \ll
\partial^2 \Phi / \partial r^2 \approx 2 M_\odot/R^3$, we then have
\begin{equation}
  \sqrt{(c_2 - 1)^3 (c_{14} + 2 c_2)}
  \ll \frac{r_\oplus^2}{M_\oplus t} \approx 1.9 \times 10^{-10}
\end{equation}
for $t \approx 5 \times 10^9$ years (the approximate age of the Earth.)
We see, therefore, that the requirement that perturbational tidal
effects remain small severely constrains our choices of $c_{14}$ and $c_2$.

\section{T\lowercase{e}V\lowercase{e}S \label{TeVeSsec}}

\subsection{Theory}

TeVeS (short for
``{\bfseries{Te}}nsor-{\bfseries{Ve}}ctor-{\bfseries{S}}calar'') is a 
modified gravity theory proposed by Bekenstein \cite{TeVeS} in an
attempt to create a fully covariant theory of Milgrom's Modified
Newtonian Dynamics (MOND).  The fields present in this theory consist
of the metric;  a vector field $u^a$ which (as in Einstein-{\ae}ther
theory) is constrained by a Lagrange multiplier $Q$ to be unit and
timelike; and two scalar fields, $\alpha$ and $\sigma$.  The
Lagrangian four-form is
\begin{equation}
\label{TevesLag}
\form{\mathcal{L}} = \left( \mathcal{L}_g + \mathcal{L}_v +
\mathcal{L}_s + \mathcal{L}_m \right) 
\form{\epsilon}
\end{equation}
where $\mathcal{L}_g$ is the usual Einstein-Hilbert action,
\begin{equation}
\label{TevesLg}
\mathcal{L}_g = \frac{1}{16 \pi} R;
\end{equation}
$\mathcal{L}_s$ is the ``scalar part'' of the action,
\begin{equation}
\label{TevesLs}
\mathcal{L}_s = - \frac{1}{2} \sigma^2 (g^{ab} - u^a u^b) \nabla_a
\alpha \nabla_b 
\alpha - \frac{1}{4} \ell^{-2} \sigma^4 F(k \sigma^2),
\end{equation}
with $k$ and $\ell$ positive constants of the theory (with ``length
dimensions'' zero and one, respectively), and $F(x)$ a free function;
$\mathcal{L}_v$ is the vector part of the action,
\begin{equation}
\label{TevesLv}
\mathcal{L}_v = - \frac{K}{32 \pi} F_{ab} F^{ab} + Q (u^a u_a + 1),
\end{equation}
with $K$ a positive dimensionless constant and $F_{ab} = \nabla_a u_b -
\nabla_b u_a$;  and $\mathcal{L}_m$ the matter action,
non-minimally coupled to the metric:
\begin{equation}
\label{TevesLm}
\mathcal{L}_m \form{\epsilon} = \form{\mathcal{L}}_\text{mat} [A, e^{2 \alpha}
  g^{ab} + 2 u^a u^b \sinh (2 \alpha) ]
\end{equation}
where $\form{\mathcal{L}}_\text{mat} [A,g^{ab}]$ would be the
minimally coupled 
matter Lagrangian for the matter fields $A$.  Note that if we ignore
the scalar and matter portions of the Lagrangian, this Lagrangian is
the same as that for Einstein-{\ae}ther theory \eqref{EAlag}, with
$c_1 = - c_3 = -K/16 \pi$ and $c_2 = c_4 = 0$.  In the present work,
we will work exclusively with the ``vacuum'' ($\mathcal{L}_m = 0$) theory.

Taking the variation of \eqref{TevesLag} to obtain the equations of
motion and the symplectic current, we find that
\begin{multline}
  \delta \form{\mathcal{L}} = \left( (\mathcal{E}_G)_{ab} \delta g^{ab}
  + (\mathcal{E}_u)_a \delta u^a + \mathcal{E}_\alpha \delta \alpha
  \right. \\ \left. +
  \mathcal{E}_\sigma \delta \sigma + \mathcal{E}_Q \delta Q +
  \nabla_a \theta^a \right) \form{\epsilon},
\end{multline}
where
\begin{subequations}
\begin{multline}
  \label{Teveseineq}
  (\mathcal{E}_G)_{ab} = \frac{1}{16 \pi} G_{ab} - \frac{1}{2} \sigma^2
  \nabla_a \alpha \nabla_b \alpha - Q u_a u_b 
  \\ + \frac{K}{16 \pi} \left( 2 u_{(a} \nabla^c F_{b)c} - F_{ac} F_b
     {}^c  + \frac{1}{4} g_{ab} F_{cd} F^{cd} \right) 
     \\ + \frac{1}{2} g_{ab} \left(
     \frac{1}{2} \sigma^2 ( \nabla^c \alpha \nabla_c \alpha - \dot{\alpha}^2
     ) + \frac{1}{4} \ell^{-2} \sigma^4 F(k \sigma^2) \right), 
\end{multline}
\begin{equation}
\label{Tevesveceq}
(\mathcal{E}_u)_a = \sigma^2 \dot{\alpha} \nabla_a \alpha +
\frac{K}{8 \pi} \nabla^b F_{ba} + 2 Q u_a,
\end{equation}
\begin{equation}
\label{Tevesaleq}
\mathcal{E}_\alpha = \nabla_a \left( \sigma^2 (\nabla^a \alpha
 - u^a \dot{\alpha} ) \right),
\end{equation}
\begin{multline}
\label{Tevessigeq}
\mathcal{E}_\sigma = -\sigma \bigg[\nabla^a 
  \alpha \nabla_a \alpha - \dot{\alpha}^2 \\ \left. + \ell^{-2}
  \sigma^2 \left( F(k 
  \sigma^2) + \frac{1}{2} k \sigma^2 F'( k \sigma^2) \right) \right],
\end{multline}
\end{subequations}
and
\begin{multline}
\label{Tevestheta}
\theta^a = \theta^a_\text{Ein} -  \sigma^2 (\nabla^a \alpha - u^a
\dot{\alpha} )  \delta \alpha \\ + \frac{K}{8 \pi} \left( F_b {}^a
\delta u^b + F^a {}_b u_c \delta g^{bc} \right).
\end{multline}
In the above, we have defined $\dot{\alpha} \equiv u^a \nabla_a
\alpha$.  The remaining equation 
$\mathcal{E}_Q = 0$ is identical to that in Einstein-{\ae}ther theory,
\eqref{EAQeq}.  For a static solution, in our usual gauge, the
background equations of motion become 
\begin{subequations}
\begin{multline}
  \label{TevesGtt0}
  (\mathcal{E}_G)_{tt} = e^{2 \Phi - 2 \Lambda} \left[ \frac{1}{16
      \pi} \left( \frac{2}{r} \fpder{\Lambda}{r} + \frac{1}{r^2} (e^{2
      \Lambda} - 1) \right) \right. \\ + \frac{K}{16 \pi}
      \left( - \fpdert{\Phi}{r} + 
    \fpder{\Phi}{r} \left( \fpder{\Lambda}{r} - \frac{2}{r} \right) -
      \frac{1}{2} \left( \fpder{\Phi}{r} \right)^2 \right) \\ \left. -
      \frac{1}{4} \sigma^2 \left( \fpder{\alpha}{r} \right)^2 -
      e^{2 \Lambda} \frac{1}{8 \ell^2} \sigma^4 F(k \sigma^2) \right],
\end{multline}
\begin{multline}
  \label{TevesGrr0}
  (\mathcal{E}_G)_{rr} = \frac{1}{16 \pi} \left(\frac{2}{r}
  \fpder{\Phi}{r} - \frac{1}{r^2} ( e^{2 \Lambda} - 1) \right) +
  \frac{K}{16 \pi} \left( \fpder{\Phi}{r} \right)^2 \\ - \frac{1}{4}
  \sigma^2 \left( \fpder{\alpha}{r} \right)^2 + e^{2 \Lambda}
  \frac{1}{8 \ell^2} \sigma^4 F(k \sigma^2) ,
\end{multline}
\begin{equation}
  \label{Tevesalph0}
  \mathcal{E}_\alpha = \frac{e^{-\Phi - \Lambda}}{r^2} \fpder{}{r}
  \left( \sigma^2 r^2 e^{\Phi - \Lambda} \fpder{\alpha}{r} \right),
\end{equation}
and
\begin{equation}
  \label{Tevessig0}
  \mathcal{E}_\sigma = - e^{-2 \Lambda} \left( \fpder{\alpha}{r}
  \right)^2 + \frac{1}{k\ell^2} y(k \sigma^2),
\end{equation}
\end{subequations}
where we have defined $y(x) = -x F(x) - \frac{1}{2} x^2 F'(x)$ (as in
\cite{TeVeS}) and used the equation for $Q$,
\begin{equation}
  \label{TevesQ0}
  Q = \frac{K}{16 \pi} e^{-2 \Lambda} \left(  - \fpdert{\Phi}{r} + 
    \fpder{\Phi}{r} \left( \fpder{\Lambda}{r} - \frac{2}{r} \right)
    \right), 
\end{equation}
to simplify.  (As in Einstein-{\ae}ther theory, the background equation
$(\mathcal{E}_u)_a = 0$ is satisfied trivially by a static {\ae}ther.)  

The symplectic form for the theory can now be calculated from
\eqref{Tevestheta}.  The result will be essentially the same as that
in Einstein-{\ae}ther theory (with the appropriate values of the
$c_i$'s), with added terms stemming from the variations of
$\mathcal{L}_s$:  
\begin{equation}
  \label{Tevesomega}
  \omega^a = \omega^a_\text{Ein} + \omega^a_\text{vec} + \omega^a_\text{s}
\end{equation}
where $\omega^a_\text{Ein}$ is given by \eqref{omegaein},
$\omega^a_\text{vec}$ is given by \eqref{EAomegavec} with $c_1 = -c_3
= - K/16 \pi$ and $c_2 = c_4 = 0$, and
\begin{multline}
  \label{Tevesomegas}
  \omega^a_\text{s} = - \sigma^2 \left[ (\nabla^a \alpha - u^a
    \dot{\alpha} ) \left( 2 \frac{\delta_1 \sigma}{\sigma} - \frac{1}{2}
    g_{bc} \delta_1 g^{bc} \right) \right. \\ +  \delta_1 g^{ab} 
    \nabla_b \alpha + (g^{ab} - u^a u^b) \nabla_b \delta_1 \alpha
    \\  - 2
    u^{(a} \delta_1 u^{b)} \nabla_b \alpha \bigg] \delta_2 \alpha.
\end{multline}

\subsection{Applying the variational formalism}

The next step is to write out the $t$-component of $\omega^a$ in terms
of the perturbational variables.  We will take our metric
perturbations to have the usual form, and our vector perturbation to
be of the same form as was used for Einstein-{\ae}ther theory
\eqref{EAupsdef}.  For the two scalar fields, we define $\delta
\alpha \equiv \beta$ and $\delta \sigma \equiv \tau$.  Calculating
$\omega^t_\text{s}$ in terms of these perturbational variables, and
using the results of \eqref{EAomegat}, we find that 
\begin{multline}
  \label{Tevesomegat}
  \omega^t = e^{-2 \Phi} \left\{ 2 \sigma^2 \fpder{\beta_1}{t} \beta_2
  \right. \\ 
  \left. + e^{\Phi} \upsilon_1 \left[ 
  \frac{K}{8 \pi} \left(\fpder{\Phi}{r} \lambda_2 - \fpder{\phi_2}{r}
  - e^{2 \Lambda - \Phi} \fpder{\upsilon_2}{t} \right) + \sigma^2
  \fpder{\alpha}{r} \beta_2 \right] 
  \right\} \\ - \switch.
\end{multline}

We now turn to the question of the constraints.  The tensor $C_{ab}$
is again given by $C_{ab} = (\mathcal{E}_G)_{ab} + u_a
(\mathcal{E}_u)_b$, and thus to first order we have $\delta C_{rt} =
(\mathcal{E}_G)_{rt} + \delta u_r (\mathcal{E}_u)_t$ (since $u^r = 0$
in the background.)  Calculating $\delta C_{rt}$ in terms of our
perturbational variables, we find that the preconstraint equation is
\begin{multline}
  \label{TevesF}
  F = r^2 e^{ \Phi - \Lambda} \left[ \sigma^2 \fpder{\alpha}{r} \beta
  - \frac{1}{8 \pi} \left( \frac{2}{r} + K \fpder{\Phi}{r} \right)
  \lambda \right. \\ \left. + \frac{K}{8 \pi} e^{2 \Lambda - \Phi}
  \fpder{\upsilon}{t} + \frac{K}{8 \pi} \fpder{\phi}{r} \right] = 0.
\end{multline}

In principle, we could now use this equation, together with the
equation
\begin{multline}
\label{TevesCrr}
0 = \delta C_{rr} = e^{2 \Lambda} \left( \frac{\sigma^4}{2 \ell^2} F(k
\sigma^2) - \frac{1}{4 \pi r^2} \right) \lambda \\+ \sigma \left(
\frac{1}{k \ell^2} e^{2 \Lambda} y(k \sigma^2) -
\left(\fpder{\alpha}{r} \right)^2 \right) \tau + e^{2 \Lambda - \Phi}
\frac{K}{8 \pi} \fpder{\Phi}{r} \fpder{\upsilon}{t} \\- \sigma^2
\fpder{\alpha}{r} \fpder{\beta}{r} + \frac{1}{8 \pi} \left(
\frac{2}{r} + K \fpder{\Phi}{r} \right) \fpder{\phi}{r} 
\end{multline}
to obtain equations for $\lambda$ and $\pder{\phi}{r}$ in terms of the
``matter'' variables $\upsilon$, $\beta$, and $\tau$.  However, it is
simpler to pursue a similar tactic to the one we used in the reduction
of Einstein-{\ae}ther theory.  To wit, the $r$-component of $(\delta
\mathcal{E}_u)_a$ is
\begin{multline}
  (\delta \mathcal{E}_u)_r = \left( 2 e^{2 \Lambda} Q + \sigma^2
  \left(\fpder{\alpha}{r} \right)^2 \right) \upsilon + e^{- \Phi}
  \sigma^2 \fpder{\alpha}{r} \fpder{\beta}{t} \\ + e^{-\Phi} \frac{K}{8
  \pi} \left( \fpder{\Phi}{r} \fpder{\lambda}{t} - e^{2 \Lambda -
  \Phi} \fpdert{\upsilon}{t} - \fpdertm{\phi} \right) = 0.
\end{multline}
We can use \eqref{TevesF} to simplify this equation;  the result
is
\begin{multline} 
  \label{Teveschievol}
  e^\Phi \left( 2 e^{2 \Lambda} Q + \sigma^2 \left(\fpder{\alpha}{r}
  \right)^2 \right) \upsilon \\ = \fpder{}{t} \left( \frac{1}{4 \pi r}
  \lambda - 2 \sigma^2 \fpder{\alpha}{r} \beta \right)
\end{multline}
We then define the new variable $\chi$ as 
\begin{equation}
  \label{Teveschidef}
  \chi = \frac{1}{4 \pi r}
  \lambda  - 2 \sigma^2 \fpder{\alpha}{r} \beta.
\end{equation} 

The evolution equation for $\beta$, meanwhile, is given by the
equation $\delta \mathcal{E}_\alpha = 0$;  in terms of the
perturbational fields, it is
\begin{multline}
  \label{Tevesbetaeq}
  - 2 e^{2 \Lambda -2 \Phi}
  \fpdert{\beta}{t} + \fpdert{\beta}{r} + \left(\frac{2}{\sigma}
  \fpder{\sigma}{r} + \fpder{\Phi}{r} - \fpder{\Lambda}{r} +
  \frac{2}{r} \right) \fpder{\beta}{r} \\ +  \fpder{\alpha}{r} \left(
  \fpder{\phi}{r} - \fpder{\lambda}{r} \right) + \frac{2}{\sigma}
  \fpder{\alpha}{r} \fpder{\tau}{r} \\ - \frac{1}{\sigma^2}
  \fpder{\alpha}{r} \fpder{\sigma}{r} \tau - e^{2 \Lambda - \Phi}
  \fpder{\alpha}{r} \fpder{\upsilon}{t} = 0.
\end{multline}
Finally, the field $\tau$, being the perturbation of the auxiliary field
$\sigma$, can be solved for algebraically in the equation $\delta
\mathcal{E}_\sigma = 0$:
\begin{equation}
  \label{Tevestau}
  \ell^{-2} \sigma y'(k \sigma^2) \tau + e^{-2 \Lambda} \left( \left(
  \fpder{\alpha}{r} \right)^2 \lambda - \fpder{\alpha}{r}
  \fpder{\beta}{r} \right) = 0.  
\end{equation}
We can thus follow the following procedure to write the evolution
equations solely in terms of $\beta$ and $\chi$:  first, we use the
preconstraint equation \eqref{TevesF} to eliminate the combination
$\pder{\phi}{r} + e^{2 \Lambda - \Phi} \pder{\upsilon}{t}$ in favour
of $\lambda$ and $\beta$;  next, we use \eqref{Tevestau}
to eliminate $\tau$ and its spatial derivatives;  and finally, we use
the definition of $\chi$ \eqref{Teveschidef} to eliminate $\lambda$
and its derivatives in favour of $\beta$ and $\chi$.  Applying our
procedure to \eqref{TevesCrr} results in an equation depending on
$\pder{\upsilon}{t}$ as well as $\beta$, $\pder{\beta}{r}$, and
$\chi$;  combining this equation with \eqref{Teveschievol} then yields
a second-order evolution equation for $\chi$ of the form
\begin{equation}
  \label{Tevesreducedchi}
  \fpdert{\chi}{t} = \mathcal{W} \chi + \mathcal{V}_3 \fpder{\beta}{r}
  + \mathcal{V}_4 \beta.
\end{equation}
where the coefficients $\mathcal{V}_i$ and $\mathcal{W}$ are dependent on
the background fields.  Similarly, applying this procedure to
\eqref{Tevesbetaeq}, and using \eqref{TevesCrr} to eliminate a
resulting $\pder{\upsilon}{t}$ term, we obtain a second-order equation
for $\beta$:
\begin{equation}
  \label{Tevesreducedbeta}
  \fpdert{\beta}{t} = \mathcal{U}_1 \fpdert{\beta}{r} +
  \mathcal{U}_2 \fpder{\beta}{r} + \mathcal{U}_3 \beta + \mathcal{V}_1
  \fpder{\chi}{r} + \mathcal{V}_2 \chi,
\end{equation}
where, again, the $\mathcal{U}_i$'s and $\mathcal{V}_i$'s depend on
the background fields.  Note that \eqref{Tevesreducedchi} has no 
dependence on the spatial derivatives of $\chi$, only upon $\chi$
itself. 

We also need to reduce the symplectic form and express it in terms of
$\beta$ and $\chi$.  Applying the preconstraint equation \eqref{TevesF}
to \eqref{Tevesomegat}, and using the definition of $\chi$
\eqref{Teveschidef} along with \eqref{Teveschievol}, we find that 
\begin{multline}
  \label{Tevessympform}
  \Omega = 4 \pi \int \dif r \, r^2 e^{\Lambda - \Phi} \left[
  \frac{1}{H} \fpder{\chi_1}{t} \chi_2  + 2 \sigma^2
  \fpder{\beta_1}{t} \beta_2 \right] \\ - \switch,
\end{multline}
where
\begin{equation}
  H = - \left( 2 e^{2 \Lambda} Q + \sigma^2 \left(\fpder{\alpha}{r} 
  \right)^2 \right).
\end{equation}
We can also apply the background equations of motion, along with the
Equation \eqref{TevesQ0}, to this quantity to obtain
\begin{equation}
  \label{TevesH2}
  H = \frac{1}{4 \pi r} \left(\fpder{\Lambda}{r} + \fpder{\Phi}{r}
  \right) - 2 \sigma^2 \left(\fpder{\alpha}{r} \right)^2.
\end{equation}

For a variational principle to exist for this theory, the
form $\form{W}_{\alpha \beta}$ defined by \eqref{Tevessympform} must
be positive definite.  In our case, this means that the coefficients
of both $(\pder{\chi_1}{t}) \chi_2$ and $(\pder{\beta_1}{t}) \beta_2$
in the integrand of \eqref{Tevessympform} must be positive for the
background solution about which we are perturbing. While the
coefficient for the latter term is obviously always positive, the
situation for the former coefficient (namely, $H$) is not so clear.
To address this issue, we need to know the properties of the
spherically symmetric static background solutions of TeVeS. These
solutions (with a ``static {\ae}ther'') are described in
\cite{Tevessph}.  In our gauge, they are most simply described in
terms of a parameter $z$:\footnote{This $z$ is the ``radial
  coordinate'' in isotropic spherical coordinates, as used in
  \cite{Tevessph}.} 
\begin{equation}
  \label{TevesBGr}
  r(z) = \frac{z^2 - z_c^2}{z} \left( \frac{
  z - z_c}{z + z_c} \right)^{- z_g / 4
  z_c}
\end{equation}
\begin{equation}
  \label{TevesBGPhi}
  e^{\Phi(z)} = \left( \frac{z - z_c}{z +
  z_c} \right)^{z_g/4 z_c}
\end{equation}
\begin{equation}
  \label{TevesBGLambda} 
  e^{\Lambda(z)} = \frac{z^2 - z_c^2}{(z^2 + z_c^2) - \frac{1}{2} z
  z_g} 
\end{equation}
\begin{equation}
  \label{TevesBGalph}
  \alpha(z) = \alpha_c + \frac{k m_s}{8 \pi z_c} \ln \left( \frac{ z -
  z_c}{z + z_c} \right)
\end{equation}
where $z_c$, $z_g$, $m_s$, and $\alpha_c$ are constants of integration.
The first three of these are related by 
\begin{equation}
  \label{Tevesrcdef}
  z_c = \frac{z_g}{4} \sqrt{ 1 + \frac{k}{\pi} \left( \frac{ m_s}{z_g}
  \right)^2 - \frac{K}{2} }
\end{equation}
while $\alpha_c$ ``sets the value of $\alpha$ at $\infty$.''  The
constant $m_s$, which can be thought of as the ``scalar charge'' of
the star, is defined by an integral over the central mass
distribution \cite{TeVeS};  for a perfect fluid with $\rho + 3 P \geq
0$, $m_s$ is non-negative.  Finally, the constant $z_g$ is 
also defined in terms of an integral over the central matter
distribution \cite{TeVeS}.

\begin{figure}
  \includegraphics[width=\figwidth]{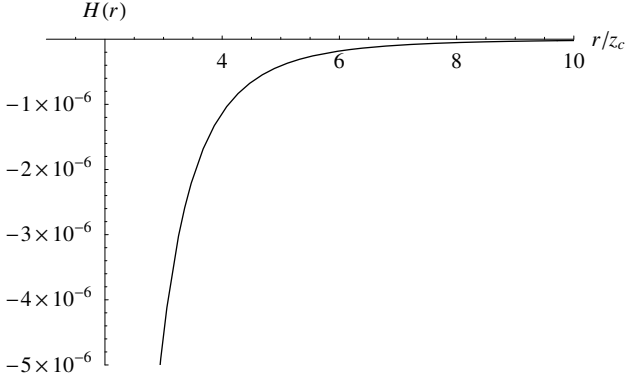}
  \caption{\label{Hfigure} Plot of $H(r)$ versus $r$ for TeVeS.  In
  this plot, $k$ and $K$ are both chosen to be $10^{-2}$, and $m_s =
  z_g = 1$.}
\end{figure}
Plotting $H(r)$ parametrically (Figure \ref{Hfigure}), we see that this
function is strictly 
negative.  In fact, in the $z \to \infty$ limit, we have $r/z = 1 +
\mathcal{O}(z_c/z)$, and so we can take $r \approx z$ to a good
approximation. Calculating $H(r)$ in terms of $\Phi(r)$ and
$\Lambda(r)$, we find that as $r \to \infty$, 
\begin{equation}
  \label{TevesHapprox}
  H(r) \approx - \frac{1}{16 \pi} \left( \frac{k}{\pi} m_s^2 +
  \frac{K}{2} z_g^2  \right) \frac{1}{r^4} 
\end{equation}
which is negative for any positive choice of $k$ and $K$.  Therefore,
it is not possible to straightforwardly apply the variational
principle to TeVeS, since the quadratic form defined by
$\form{W}_{\alpha \beta}$ is indefinite and thus cannot used as an
inner product.\footnote{ Note  that this also implies that the
  ``perturbational Hamiltonian'' of TeVeS, as defined in equation (54)
  of \cite{GVP}, has an indefinite kinetic term.}

\subsection{WKB analysis}

While we cannot derive a variational principle with TeVeS, the fact
that we have been able to ``reduce'' the equations of motion to an
unconstrained form still allows us to analyse the stability of its
spherically symmetric solutions.  Let us consider a WKB ansatz, of the
form  
\begin{equation}
  \label{Tevesansatz}
  \begin{bmatrix} \beta(r) \\ \chi(r) \end{bmatrix} = e^{i(\omega(r) t +
    \kappa r)} 
  \begin{bmatrix} f_\beta(r) \\ f_\chi(r) \end{bmatrix}
\end{equation}
with $\kappa$ very large compared to the scale of variation of the
background functions $\mathcal{U}_i$, $\mathcal{V}_i$, and
$\mathcal{W}$.  We will further choose the the functions $f_i(r)$ and
$\omega(r)$ are chosen to be ``slowly varying'' relative to the scale
defined by $\kappa$, i.e. 
\begin{align}
  \label{Tevesslowvary}
  \frac{1}{f_i(r)} \fpder{f_i}{r} &\ll \kappa, &
  \frac{1}{\omega(r)} \fpder{\omega}{r} &\ll \kappa.
\end{align}
Under this assumption, we will then have 
\begin{equation}
  \fpder{}{r} \left( e^{i(\omega(r) t + \kappa r)} f_i(r) \right)
  \approx i \kappa e^{i(\omega(r) t + \kappa r)} f_i(r). 
\end{equation}

Now let us apply the time-evolution operator $\mathcal{T}$ implicitly
defined by \eqref{Tevesreducedbeta} and \eqref{Tevesreducedchi} to our
ansatz \eqref{Tevesansatz}.  We see that for sufficiently large
$\kappa$, the highest-derivative terms will dominate the
lower-derivative terms.  Thus, to a good approximation we will have
\begin{equation}
  \mathcal{T} \begin{bmatrix} \beta \\ \chi \end{bmatrix} \approx
  \begin{bmatrix} -\kappa^2 \mathcal{U}_1(r) & i \kappa
    \mathcal{V}_1(r) \\ i \kappa \mathcal{V}_3 (r) &
    \mathcal{W}(r) \end{bmatrix} 
  \begin{bmatrix} f_\beta(r) \\ f_\chi(r) \end{bmatrix} e^{i(\omega(r) t +
    \kappa r)}.
\end{equation}
Then, in the limit of large $\kappa$, our ansatz \eqref{Tevesansatz}
will be an approximate eigenvector  of $\mathcal{T}$ if there exist a
$f_\beta(r)$ and $f_\chi (r)$ such that 
\begin{equation}
  \label{Teveskmatrix}
  - \omega^2(r) \begin{bmatrix} f_\beta(r) \\ f_\chi(r) \end{bmatrix} =
  \begin{bmatrix} -\kappa^2 \mathcal{U}_1(r) & i \kappa
    \mathcal{V}_1(r) \\ i \kappa \mathcal{V}_3 (r) &
    \mathcal{W}(r) \end{bmatrix} 
  \begin{bmatrix} f_\beta(r) \\ f_\chi(r) \end{bmatrix}.
\end{equation}
In other words, in the limit of large $\kappa$, the problem of finding
modes of $\mathcal{T}$ is a simple two-dimensional eigenvalue problem
where the eigenvalues are functions of $r$.  In this limit, the
eigenvalues of this matrix are (to leading order in $\kappa$)
\begin{equation}
  \omega^2(r) \approx \left\{ \kappa^2 \mathcal{U}_1(r), -
  \mathcal{W}(r) + \frac{ \mathcal{V}_1 (r) \mathcal{V}_3
  (r)}{\mathcal{U}_1 (r)} \right\}.
\end{equation}
It is not difficult to verify that for sufficiently large $\kappa$,
our assumptions for the ansatz \eqref{Tevesslowvary} are satisfied.

It remains to write out the functions $\mathcal{U}_1(r)$,
$\mathcal{V}_1(r)$, $\mathcal{V}_3(r)$, and $\mathcal{W}(r)$ in terms
of the background functions.  These can be shown to be:
\begin{equation}
  \mathcal{U}_1 = \frac{1}{2} e^{2 \Phi - 2 \Lambda} \left( 1 +
  e^{-2 \Lambda} \frac{2 \ell^2 (\fpder{\alpha}{r})^2}{ \sigma^2 y'(k
  \sigma^2)} 
  \right) ,
\end{equation}
\begin{equation}
  \mathcal{V}_1 = -\frac{r}{4} e^{2 \Phi - 2 \Lambda}
  \fpder{\alpha}{r} \left( 1 + e^{-2 \Lambda} 
  \frac{2 \ell^2 (\fpder{\alpha}{r})^2}{ \sigma^2 y'(k \sigma^2)}
  \right) ,
\end{equation}
\begin{multline}
  \mathcal{V}_3 = 16 \pi r e^{2 \Phi - 2 \Lambda} \sigma^2
  \fpder{\alpha}{r} \\ \times \left( 8 \pi \sigma^2 \left( \fpder{\alpha}{r}
  \right) - \frac{1}{r} \left( \fpder{\Lambda}{r} + \fpder{\Phi}{r}
  \right) \right),
\end{multline}
and
\begin{multline}
  \mathcal{W} = r^2 e^{2 \Phi - 2 \Lambda} \left( \frac{2}{K} - 1 - 8 \pi
  \sigma^2 \left(\fpder{\alpha}{r} \right)^2 \right) \\ \times \left( 8 \pi
  \sigma^2 \left(\fpder{\alpha}{r} \right)^2 - \frac{1}{r} \left(
  \fpder{\Lambda}{r} + \fpder{\Phi}{r}  \right) \right) .
\end{multline}
This implies that the eigenvalues of the matrix in
\eqref{Teveskmatrix} are  
\begin{equation}
  \label{Tevesomega1}
  \omega^2(r) \approx \frac{\kappa^2}{2} e^{2 \Phi - 2 \Lambda} \left(
  1 + e^{-2 \Lambda} \frac{2 \ell^2 (\fpder{\alpha}{r})^2}{ \sigma^2
  y'(k \sigma^2)} 
  \right)
\end{equation}
and
\begin{multline}
  \label{Tevesomega2}
  \omega^2(r) \approx
    e^{2 \Phi - 2 \Lambda} \left( \frac{2}{K} - 1 \right) \\
    \times \left( \frac{1}{r} \left( \fpder{\Lambda}{r} +
    \fpder{\Phi}{r} \right) - 
    8 \pi \sigma^2 \left( \fpder{\alpha}{r} \right)^2
    \right).
\end{multline}

We can see that this first eigenvalue \eqref{Tevesomega1} is always
positive as long as $y'(x) > 0$;  indeed, the choice of $y(x)$ made in 
\cite{TeVeS} does satisfy this inequality.  The second eigenvalue
\eqref{Tevesomega2}, however, is just
\begin{equation}
  \omega^2(r) = 4 \pi e^{2 \Phi - 2 \Lambda} \left(\frac{2}{K}
  - 1 \right) H(r).
\end{equation}
Since $H(r)$ is always negative, we conclude that this second mode is
unstable for $0 < K < 2$.  Further, for a spherically symmetric
solution outside a Newtonian star, the approximation
\eqref{TevesHapprox} is valid;  thus, to lowest non-vanishing order in
$r^{-1}$, we have 
\begin{equation}
  \omega^2(r) \approx - \frac{1}{4} \left( \frac{2}{K} - 1 \right) \left(
  \frac{k}{\pi} m_s^2+ \frac{K}{2} z_g^2 \right) \frac{1}{r^4}.
\end{equation}

\subsection{Discussion}

The above result for $\omega^2(r)$ implies that for stability of
Newtonian solutions, we must have $K > 2$, ruling out the range of
parameters originally considered 
in \cite{TeVeS}.  To estimate the time scale of the instability when 
$K < 2$, we first note that unless $k$ and $K$ are much larger than
the ratio of the star's radius $R$ to its Schwarzchild radius $m_g$, the
parameters $m_g$, $m_s$, and $z_g$ are all approximately equal,
differing by terms of $\mathcal{O}(m_g/R)$ \cite{TeVeS}.  In
particular, if $k$ and $K$ are of the same order and $K \not \approx 2$,
we find that the timescale of this instability will be on the order of
$10^6$ seconds---approximately two weeks---for points near the surface
of the Sun.  

Note also that the effective gravitational potential for
nonrelativistic motion is 
\begin{equation}
  \Phi = \Xi \Phi_N + \alpha,
\end{equation}
where 
\begin{equation}
  \Xi = e^{-2 \alpha_c} (1 - K/2)^{-1},
\end{equation}
$\Phi_N$ is the Newtonian gravitational potential, and $\alpha_c$ is
the asymptotic value of $\alpha$, determined by the cosmological
boundary conditions \cite{BekReview}.  Expanding the asymptotic
behaviour of $\Phi_N$ and $\alpha$ about infinity, we find that
\begin{equation}
  \label{Teveseffpot}
  \Phi \approx \alpha_c - e^{-2 \alpha_c} \left( \frac{1}{1 - K/2} +
  \frac{k}{4 \pi} \right) \frac{m_g}{r} 
\end{equation}
where in the weak-field limit, we have
\begin{equation}
  m_g \approx 4 \pi \int r^2 \rho
\end{equation}
and $m_s \approx e^{-2 \alpha_c} m_g$.  The coefficient in parentheses in
\eqref{Teveseffpot} can be thought of as renormalizing the gravitational
constant $G$ by some overall factor;  the first term in parentheses
comes from the usual perturbations to the ``physical metric''
component $\tilde{g}_{tt}$, while the second term comes from variation
of the scalar field $\alpha$.  However, if $K > 2$, this would
make the metric contribution to the ``effective gravitational
constant'' negative.  This could be remedied by choosing $k > 8 \pi /
(K - 2)$.  However, it is unclear whether the theory, in this
parameter regime, would still be experimentally viable;  for example,
Big Bang nucleosynthesis constraints require that $k < 0.75$, and
the CMB power spectrum also constrains the viable regions
of parameter space \cite{Skor}.\footnote{We
  also note that if the gravitational effects of the scalar
  are dominant over those of the tensor, the solutions of TeVeS would
  greatly resemble those of the so-called ``stratified theories'' with
  time-orthogonal space slices \cite{Will}.  As these are ruled out
  experimentally via geophysical experiments, we could not make $k$ or
  $K$ too large without running afoul of these experimental
  constraints.} 

We also note that the eigenvector corresponding to the unstable
mode of $\mathcal{T}$ will satisfy (to leading order)
\begin{equation}
  i \kappa f_\beta + \frac{\mathcal{V}_1}{\mathcal{U}_1} f_\chi
  \approx 0
\end{equation}
or, in the limit of large $\kappa$, $f_\beta \approx 0$.  In other
words, this unstable mode should manifest itself in growth of the
radial component of the vector field $u^a$ rather than growth of the
scalar $\alpha$  (cf.\ \eqref{Teveschievol}).  An instability of the
vector field was also found in \cite{TeVeScosmo1} by Dodelson and
Liguori.  However, it seems unlikely that this is the same instability
for three reasons.  First, the instability found in \cite{TeVeScosmo1}
was found in a cosmological context, not a Newtonian-gravity context;
since the cosmological solutions of TeVeS are in a very real sense
separate from the Newtonian solutions (existing on two different
branches of the function $y(x)$), it is difficult to draw a direct
correspondence between the stability properties of these two types of
solutions.  Second, the instability found in \cite{TeVeScosmo1}
manifests itself only in the limit of a matter-dominated Universe;
the instability we have found exists \textsl{in vacuo}.  Third,
Dodelson and Liguori's instability requires $K$ to be sufficiently
small relative to $k$, while our instability is present for all $k >
0$ and $0 < K < 2$.  

Finally, it is perhaps notable that the vector field $u^a$ in TeVeS has
``Maxwellian'' kinetic terms, which we found in Section
\ref{aethersec} to be unstable in the context of Einstein-{\ae}ther
theory.  It is possible, then, that this instability could be cured
via a more general kinetic term for the vector field.  

\appendix

\begin{widetext}

\section{Coefficients for the $f(R)$ variational principle \label{fRapp}}

The quantity $(\psi, \mathcal{T} \psi)$ for
$f(R)$ gravity can be put in the form 
\begin{equation}
  (\psi, \mathcal{T} \psi) = 4 \pi \int \dif r \, \left[ \mathcal{C}_1
    \left( \fpder{b}{r} \right)^2  + \mathcal{C}_2 \left( \fpder{\zeta}{r}
    \right)^2 + \mathcal{C}_3 \left( \zeta \fpder{b}{r}
    - b \fpder{\zeta}{r} \right)  + \mathcal{C}_4 b^2  + \mathcal{C}_5
    \zeta^2 + 2 \mathcal{C}_6 b \zeta \right]
\end{equation}
where the $\mathcal{C}_i$ coefficients are given by
\begin{equation}
  \mathcal{C}_1 = 6 e^{\Phi - \Lambda} \mathcal{S}^{-2} f'(\alpha),
\end{equation}
\begin{equation}
  \mathcal{C}_2 = 8 \pi r^2 e^{\Lambda + \Phi} \varrho'' \nu^2 ,
\end{equation}
\begin{equation}
  \mathcal{C}_3 = 16 \pi r^2 e^{\Phi + \Lambda} \mathcal{S}^{-1}
  \left( \frac{3}{r^2}  \mathcal{S}^{-1} f'(\alpha) ( \varrho'
  \nu - \varrho'' \nu^2) -  \varrho' \nu \left( \fpder{\Phi}{r} +
  \frac{2}{r} \right) \right),
\end{equation}
\begin{multline}
  \mathcal{C}_4 = r e^{\Phi - \Lambda} \Bigg\{ \frac{6}{r}
  \mathcal{S}^{-1} \left( 2 \fpder{ \Lambda}{r} - \fpder{\Phi}{r} +
  \frac{2}{r} \right) \\ \qquad + \mathcal{S}^{-2} \Bigg[ \frac{2}{r}
    e^{2 \Lambda} \frac{(f'(\alpha))^2}{f''(\alpha)} - e^{2 \Lambda}
    \left(\frac{48 \pi}{r} \varrho' \nu + (f(\alpha) - 16 \pi \varrho)
    \left( \fpder{\Lambda}{r} - \fpder{\Phi}{r} + \frac{3}{r} \right)
    \right) + 2 f'(\alpha) \left( e^{2 \Lambda} \alpha \left(
    \fpder{\Lambda}{r} - \fpder{\Phi}{r} + \frac{5}{r} \right) \right. \\
    \left.- \frac{3}{r} \left(\frac{1}{r^2} (7 e^{2 \Lambda} - 1) + 3
    \left( \fpder{\Lambda}{r} \right)^2 + \fpder{\Lambda}{r} \left(
    \frac{1}{r} (e^{2 \Lambda} + 5) - 3 \fpder{\Phi}{r} \right) -
    \frac{1}{r} \fpder{\Phi}{r} (e^{2 \Lambda} + 7) - \frac{\partial^2
      \Lambda}{\partial r^2} \right) \right) \Bigg] \\ \qquad + \frac{6}{r}
  \mathcal{S}^{-3} f'(\alpha) \left[ - 8 \pi e^{2 \Lambda}
    \fpder{\Phi}{r} \frac{(\varrho')^2}{\varrho''} + \frac{2}{r} e^{2
      \Lambda} (f(\alpha) - 16 \pi \varrho) + 2 f'(\alpha) \left( - e^{2
      \Lambda} \alpha + \frac{3}{r}\left( \fpder{\Lambda}{r} -
    \fpder{\Phi}{r} + \frac{1}{r} (e^{2 \Lambda} - 1) \right) \right)
    \right.  \\ \left.   + 8 \pi e^{2 \Lambda} \varrho' \nu \left(
    \fpder{\Lambda}{r} + \fpder{\Phi}{r} + \frac{8}{r} \right) \right] \\
  - \mathcal{S}^{-4} \frac{12}{r} e^{2 \Lambda} f'(\alpha) \left( e^{2
    \Lambda} \left(8 \pi \varrho' \nu \right)^2 + \frac{24 \pi}{r^2}
  f'(\alpha) \left( 3 \varrho' \nu - \varrho'' \nu^2 \right) \right)
  \Bigg\},
\end{multline}
\begin{multline}
  \mathcal{C}_5 = 8 \pi r^2 e^{\Lambda + \Phi} \mathcal{S}^{-2} \left\{
  16 \pi e^{2\Lambda} \varrho' \nu \left( 8 \pi e^{2 \Lambda} \varrho''
  \varrho'  \nu^3 + \frac{3}{r^2} f'(\alpha) \left( \varrho'' \nu^2 +
  \varrho' \nu \right) \right) \right. \\ + 8 \pi e^{2 \Lambda}
  \mathcal{S} \varrho' \nu \left(  \varrho'' \nu^2 \left(
  \fpder{\Phi}{r} - \frac{4}{r} \right) + \varrho' \nu \left(
  \left(\frac{\varrho''' \nu}{\varrho''}  - 1 \right)\fpder{\Phi}{r} -
  \frac{4}{r} \right) \right) \\ \left. - \varrho' \nu  \left( \left(
  \fpder{\Phi}{r} \right)^2 + 2 \fpder{\Lambda}{r} \left(
  \fpder{\Phi}{r} + \frac{1}{r} \right) + \frac{1}{r^2} \left( e^{2
    \Lambda} - 1 \right) - e^{2 \Lambda} \frac{\alpha}{2} \right) +
  \frac{\varrho''' \nu}{\varrho''} \fpder{\Phi}{r} \left(
  \fpder{\Lambda}{r} + \frac{2}{r} \right) \right\},
\end{multline}
and
\begin{multline}
  \mathcal{C}_6 = 8 \pi e^{\Lambda + \Phi} r^2 \mathcal{S}^{-3} \left\{
  \mathcal{S}^2 \left[ \frac{(\varrho')^2}{\varrho''} \fpder{\Phi}{r}
    \left( \fpder{\Phi}{r} + \frac{2}{r} \right) + \varrho' \nu \left(
    \frac{1}{r^2} + \frac{4}{r} \fpder{\Phi}{r} + \left( \fpder{\Phi}{r}
    \right)^2 + \fpder{\Lambda}{r} \left( \fpder{\Phi}{r} + \frac{2}{r}
    \right) \right) - \frac{3}{r^2} \varrho'' \nu^2 \right] \right. \\  +
  \frac{12}{r^2} f'(\alpha) \left( 8 \pi \varrho'' \varrho' \nu^3 +
  \frac{3}{r^2} \left( \varrho' \nu + \varrho'' \nu^2 \right) \right) \\
  - \mathcal{S} \left[ e^{2 \Lambda} \varrho' \nu \left( \frac{1}{r}
    (f(\alpha) - 16 \pi \varrho) + 8 \pi \varrho' \nu \left(
    \fpder{\Phi}{r} + \frac{2}{r} \right) \right) + \frac{1}{r} f'(\alpha)
    \left( \frac{3}{r} \fpder{\Phi}{r} \frac{(\varrho')^2}{\varrho''} -
    \frac{3}{r} \varrho'' \nu^2 \left( 3 \fpder{\Lambda}{r} +
    \fpder{\Phi}{r} + \frac{10}{r} \right) \right. \right. \\
    \left. \left. \left. + \varrho' \nu \left( \frac{3}{r} \left(
    \frac{2}{r} (e^{2 \Lambda} - 4) + \fpder{\Lambda}{r} + \fpder{\Phi}{r}
    \left( \frac{\varrho''' \nu}{\varrho''} - 2 \right) \right) - 2 e^{2
      \Lambda} \alpha \right) \right) \right] \right\}.
\end{multline}

\section{Symplectic current for Einstein-{\ae}ther theory \label{EAapp}}

The symplectic current for vacuum Einstein-{\ae}ther theory is given by
$\omega^a = \omega^a_\text{Ein} + \omega^a_\text{vec}$, where
$\omega^a_\text{Ein}$ is the standard Einstein symplectic form (given
by \eqref{omegaein}), and  
\begin{multline}
  \label{EAomegavec}
  \omega^a_\text{vec} = (M_{(1)})^a {}_{bc} {}^d {}_{ef} \delta_2
  g^{bc} \nabla_d \delta_1 g^{ef} + (M_{(2)})^a {}_{bcde} \delta_2 g^{bc}
  \delta_1 g^{de} + (M_{(3)})^a {}_{bc} {}^d {}_e \delta_2 g^{bc}
  \nabla_d \delta_1 u^e + (M_{(4)})^a {}_{bcd} \delta_2 g^{bc} \delta_1
  u^d \\ + (M_{(5)})^a {}_b {}^c {}_{de} \delta_2 u^b \nabla_c \delta_1
  g^{de} + (M_{(6)})^a {}_b {}^c {}_d \delta_2 u^b \nabla_c \delta_1 u^d
  + (M_{(7)})^a {}_{bc} \delta_2 u^b \delta_1 u^c - \switch,
\end{multline}
where $M_{(i)}$ tensors are given by 
\begin{multline}
(M_{(1)})^a {}_{bc} {}^d {}_{ef} = c_1 \left( g^{ad} g_{be} u_c u_f -
  \delta^a {}_e \delta^d {}_b u_c u_f + \frac{1}{2} u^a u^d g_{be}
  g_{cf} \right) + \frac{1}{2} c_2 u^a u^d g_{bc} g_{ef} \\ + c_3 \left(
  \delta^a {}_e \delta^d {}_b u_c u_f - g^{ad} g_{be} u_c u_f +
  \frac{1}{2} u^a u^d g_{be} g_{cf} \right) - c_4 \left( 2 u^a u^d
  g_{be} - \delta^a {}_e u^d u_b - u^a u_e \delta^d {}_b
   + \frac{1}{2} g^{ad} u_b u_e \right) u_c u_f,
\end{multline}
\begin{multline}
(M_{(2)})^a {}_{bcde} = (c_3 - c_1) \left( \nabla_b u^a g_{ce} u_d +
  \delta^a {}_d u_c \nabla_e u_b + u^a g_{bd} \nabla_c u_e  + \frac{1}{2} g_{de} u_c 
  ( \nabla_b u^a - \nabla^a u_b) \right)   \\
  + \frac{1}{2} (c_1 + c_3) u^a \nabla_b u_c g_{de} - c_4 \left( u^a \dot{u}_c
   - \frac{1}{2} \dot{u}^a u_c \right) u_b g_{de},
\end{multline}
\begin{multline}
(M_{(3)})^a {}_{bc} {}^d {}_e = c_1 \left( \delta^a {}_e \delta^d {}_b
  u_c - g^{ad} g_{be} u_c - u^a \delta^d {}_b g_{ce} \right) - c_2 u^a
  g_{bc} \delta^d {}_e \\ + c_3 \left( g^{ad} g_{be} u_c - \delta^a {}_e
  \delta^d {}_b u_c - u^a \delta^d {}_b g_{ce} \right) - c_4 \left( \delta^a
  {}_e u_c - 2 u^a g_{ce} \right) u_b u_d,
\end{multline}
\begin{multline}
(M_{(4)})^a {}_{bcd} = c_1 \left( \nabla_b u^a g_{cd} + \nabla^a u_b
  g_{cd} + \nabla^a u_d g_{bc} - \delta^a {}_d \nabla_b u_c - 2 \delta^a
  {}_c \nabla_b u_d \right) \\ + c_3 \left( \nabla^a u_b g_{cd} -
  \nabla_b u^a g_{cd} - \delta^a {}_d \nabla_c u_b + \nabla_d u^a
  g_{bc} \right) \\ - c_4 \left( 2 \dot{u}^a u_b g_{cd} + u_b u_c
  \nabla_d u^a - 2 \delta^a {}_d \dot{u}_b u_c - 2 u^a u_b \nabla_d
  u_c + u^a g_{bc} \dot{u}_d \right),
\end{multline}
\begin{equation}
(M_{(5)})^a {}_b {}^c {}_{de} = K^a {}_{db} {}^c u_e - K^{ac} {}_{bd} u_e
  - K^a {}_{dbe} u^c,
\end{equation}
\begin{equation}
(M_{(6)})^a {}_b {}^c {}_d = 2 K^{ac} {}_{bd},
\end{equation}
and
\begin{equation}
(M_{(7)})^a {}_{bc} = - 2 c_4 (\delta^a {}_c \dot{u}_b + u^a \nabla_c
  u_b ).
\end{equation}

\pagebreak
\end{widetext}

\begin{acknowledgments}
  This material is based upon work supported under NSF Grant
  Nos.\ DGE-0202337, DGE-0638477, and PHY-0456619.  I would like to
  thank the Chicago chapter of the ARCS Foundation for their generous
  support.  Finally, I would like to thank Jacob Bekenstein, Bob
  Geroch, Stefan Hollands, Iggy Sawicki, and above all Bob Wald for
  feedback and constructive criticism throughout.
\end{acknowledgments}

\bibliography{varprinapps}

\begin{thebibliography}{32}
\expandafter\ifx\csname natexlab\endcsname\relax\def\natexlab#1{#1}\fi
\expandafter\ifx\csname bibnamefont\endcsname\relax
  \def\bibnamefont#1{#1}\fi
\expandafter\ifx\csname bibfnamefont\endcsname\relax
  \def\bibfnamefont#1{#1}\fi
\expandafter\ifx\csname citenamefont\endcsname\relax
  \def\citenamefont#1{#1}\fi
\expandafter\ifx\csname url\endcsname\relax
  \def\url#1{\texttt{#1}}\fi
\expandafter\ifx\csname urlprefix\endcsname\relax\def\urlprefix{URL }\fi
\providecommand{\bibinfo}[2]{#2}
\providecommand{\eprint}[2][]{\url{#2}}

\bibitem[{\citenamefont{Seifert and Wald}()}]{GVP}
\bibinfo{author}{\bibfnamefont{M.~D.} \bibnamefont{Seifert}} \bibnamefont{and}
  \bibinfo{author}{\bibfnamefont{R.~M.} \bibnamefont{Wald}},
  \bibinfo{note}{preprint: \ttfamily gr-qc/0612121\rmfamily}.

\bibitem[{\citenamefont{Wald}(1991)}]{Waldstability}
\bibinfo{author}{\bibfnamefont{R.~M.} \bibnamefont{Wald}},
  \bibinfo{journal}{J.\ Math.\ Phys.} \textbf{\bibinfo{volume}{33}},
  \bibinfo{pages}{248} (\bibinfo{year}{1991}).

\bibitem[{\citenamefont{Chandrasekhar}(1964{\natexlab{a}})}]{Chandra1}
\bibinfo{author}{\bibfnamefont{S.}~\bibnamefont{Chandrasekhar}},
  \bibinfo{journal}{Astrophys. J.} \textbf{\bibinfo{volume}{140}},
  \bibinfo{pages}{417} (\bibinfo{year}{1964}{\natexlab{a}}).

\bibitem[{\citenamefont{Chandrasekhar}(1964{\natexlab{b}})}]{Chandra2}
\bibinfo{author}{\bibfnamefont{S.}~\bibnamefont{Chandrasekhar}},
  \bibinfo{journal}{Phys. Rev. Lett.} \textbf{\bibinfo{volume}{12}},
  \bibinfo{pages}{114} (\bibinfo{year}{1964}{\natexlab{b}}).

\bibitem[{\citenamefont{Carroll et~al.}(2004)\citenamefont{Carroll, Duvvuri,
  Trodden, and Turner}}]{CDTT}
\bibinfo{author}{\bibfnamefont{S.~M.} \bibnamefont{Carroll}},
  \bibinfo{author}{\bibfnamefont{V.}~\bibnamefont{Duvvuri}},
  \bibinfo{author}{\bibfnamefont{M.}~\bibnamefont{Trodden}}, \bibnamefont{and}
  \bibinfo{author}{\bibfnamefont{M.~S.} \bibnamefont{Turner}},
  \bibinfo{journal}{Phys.\ Rev.} \textbf{\bibinfo{volume}{D70}},
  \bibinfo{pages}{043528} (\bibinfo{year}{2004}).

\bibitem[{\citenamefont{Jacobson and Mattingly}(2001)}]{Aether1}
\bibinfo{author}{\bibfnamefont{T.}~\bibnamefont{Jacobson}} \bibnamefont{and}
  \bibinfo{author}{\bibfnamefont{D.}~\bibnamefont{Mattingly}},
  \bibinfo{journal}{Phys.\ Rev.} \textbf{\bibinfo{volume}{D64}},
  \bibinfo{pages}{024028} (\bibinfo{year}{2001}).

\bibitem[{\citenamefont{Jacobson and Mattingly}(2004)}]{Aetherwave}
\bibinfo{author}{\bibfnamefont{T.}~\bibnamefont{Jacobson}} \bibnamefont{and}
  \bibinfo{author}{\bibfnamefont{D.}~\bibnamefont{Mattingly}},
  \bibinfo{journal}{Phys. Rev.} \textbf{\bibinfo{volume}{D70}},
  \bibinfo{pages}{024003} (\bibinfo{year}{2004}).

\bibitem[{\citenamefont{Bekenstein}(2004)}]{TeVeS}
\bibinfo{author}{\bibfnamefont{J.~D.} \bibnamefont{Bekenstein}},
  \bibinfo{journal}{Phys. Rev.} \textbf{\bibinfo{volume}{D70}},
  \bibinfo{pages}{083509} (\bibinfo{year}{2004}).

\bibitem[{\citenamefont{Wald}(1984)}]{WaldGR}
\bibinfo{author}{\bibfnamefont{R.~M.} \bibnamefont{Wald}},
  \emph{\bibinfo{title}{General Relativity}} (\bibinfo{publisher}{University of
  Chicago Press}, \bibinfo{year}{1984}).

\bibitem[{\citenamefont{Misner et~al.}(1973)\citenamefont{Misner, Thorne, and
  Wheeler}}]{MTW}
\bibinfo{author}{\bibfnamefont{C.~W.} \bibnamefont{Misner}},
  \bibinfo{author}{\bibfnamefont{K.~S.} \bibnamefont{Thorne}},
  \bibnamefont{and} \bibinfo{author}{\bibfnamefont{J.~A.}
  \bibnamefont{Wheeler}}, \emph{\bibinfo{title}{Gravitation}}
  (\bibinfo{publisher}{W.\ H.\ Freeman}, \bibinfo{year}{1973}),
  chap.~\bibinfo{chapter}{23}, pp. \bibinfo{pages}{616--7}.

\bibitem[{\citenamefont{Nojiri and Odintsov}(2003)}]{Odintsov}
\bibinfo{author}{\bibfnamefont{S.}~\bibnamefont{Nojiri}} \bibnamefont{and}
  \bibinfo{author}{\bibfnamefont{S.~D.} \bibnamefont{Odintsov}},
  \bibinfo{journal}{Phys.\ Rev.} \textbf{\bibinfo{volume}{D68}},
  \bibinfo{pages}{123512} (\bibinfo{year}{2003}).

\bibitem[{\citenamefont{Chiba}(2003)}]{Chiba}
\bibinfo{author}{\bibfnamefont{T.}~\bibnamefont{Chiba}},
  \bibinfo{journal}{Phys.\ Lett.} \textbf{\bibinfo{volume}{B575}},
  \bibinfo{pages}{1} (\bibinfo{year}{2003}).

\bibitem[{\citenamefont{Whitt}(1984)}]{Whitt}
\bibinfo{author}{\bibfnamefont{B.}~\bibnamefont{Whitt}},
  \bibinfo{journal}{Phys.\ Lett.} \textbf{\bibinfo{volume}{B145}},
  \bibinfo{pages}{176} (\bibinfo{year}{1984}).

\bibitem[{\citenamefont{Erickcek et~al.}(2006)\citenamefont{Erickcek, Smith,
  and Kamionkowski}}]{fRint1}
\bibinfo{author}{\bibfnamefont{A.~L.} \bibnamefont{Erickcek}},
  \bibinfo{author}{\bibfnamefont{T.~L.} \bibnamefont{Smith}}, \bibnamefont{and}
  \bibinfo{author}{\bibfnamefont{M.}~\bibnamefont{Kamionkowski}},
  \bibinfo{journal}{Phys. Rev.} \textbf{\bibinfo{volume}{D74}},
  \bibinfo{pages}{121501} (\bibinfo{year}{2006}).

\bibitem[{\citenamefont{Chiba et~al.}(2006)\citenamefont{Chiba, Smith, and
  Erickcek}}]{fRint2}
\bibinfo{author}{\bibfnamefont{T.}~\bibnamefont{Chiba}},
  \bibinfo{author}{\bibfnamefont{T.~L.} \bibnamefont{Smith}}, \bibnamefont{and}
  \bibinfo{author}{\bibfnamefont{A.~L.} \bibnamefont{Erickcek}}
  (\bibinfo{year}{2006}), \bibinfo{note}{preprint: \ttfamily
  astro-ph/0611867\rmfamily}.

\bibitem[{\citenamefont{Multam{\"a}ki and Vilja}(2006)}]{fRint3}
\bibinfo{author}{\bibfnamefont{T.}~\bibnamefont{Multam{\"a}ki}}
  \bibnamefont{and} \bibinfo{author}{\bibfnamefont{I.}~\bibnamefont{Vilja}}
  (\bibinfo{year}{2006}), \bibinfo{note}{preprint: \ttfamily
  astro-ph/0612775\rmfamily}.

\bibitem[{\citenamefont{Zhang}(2007)}]{fRrefute}
\bibinfo{author}{\bibfnamefont{P.}~\bibnamefont{Zhang}} (\bibinfo{year}{2007}),
  \bibinfo{note}{preprint: \ttfamily astro-ph/0701662\rmfamily}.

\bibitem[{\citenamefont{Hu and Sawicki}(2007)}]{fRsolar}
\bibinfo{author}{\bibfnamefont{W.}~\bibnamefont{Hu}} \bibnamefont{and}
  \bibinfo{author}{\bibfnamefont{I.}~\bibnamefont{Sawicki}}
  (\bibinfo{year}{2007}), \bibinfo{note}{preprint: \ttfamily 0705.1158
  \rmfamily [astro.ph]}.

\bibitem[{\citenamefont{Dolgov and Kawasaki}(2003)}]{DolKaw}
\bibinfo{author}{\bibfnamefont{A.~D.} \bibnamefont{Dolgov}} \bibnamefont{and}
  \bibinfo{author}{\bibfnamefont{M.}~\bibnamefont{Kawasaki}},
  \bibinfo{journal}{Phys.\ Lett.} \textbf{\bibinfo{volume}{B573}},
  \bibinfo{pages}{1} (\bibinfo{year}{2003}).

\bibitem[{\citenamefont{Harada}(1997)}]{Harada}
\bibinfo{author}{\bibfnamefont{T.}~\bibnamefont{Harada}},
  \bibinfo{journal}{Prog.\ Theo.\ Phys.} \textbf{\bibinfo{volume}{98}},
  \bibinfo{pages}{359} (\bibinfo{year}{1997}).

\bibitem[{\citenamefont{Burnett and Wald}(1990)}]{BurnettWald}
\bibinfo{author}{\bibfnamefont{G.~A.} \bibnamefont{Burnett}} \bibnamefont{and}
  \bibinfo{author}{\bibfnamefont{R.~M.} \bibnamefont{Wald}},
  \bibinfo{journal}{Proc. R. Soc. Lond.} \textbf{\bibinfo{volume}{A430}},
  \bibinfo{pages}{57} (\bibinfo{year}{1990}).

\bibitem[{\citenamefont{Zhang}(2006)}]{Zhang}
\bibinfo{author}{\bibfnamefont{P.}~\bibnamefont{Zhang}},
  \bibinfo{journal}{Phys. Rev.} \textbf{\bibinfo{volume}{D73}},
  \bibinfo{pages}{123504} (\bibinfo{year}{2006}).

\bibitem[{\citenamefont{Song et~al.}(2007)\citenamefont{Song, Hu, and
  Sawicki}}]{SoHuSa}
\bibinfo{author}{\bibfnamefont{Y.-S.} \bibnamefont{Song}},
  \bibinfo{author}{\bibfnamefont{W.}~\bibnamefont{Hu}}, \bibnamefont{and}
  \bibinfo{author}{\bibfnamefont{I.}~\bibnamefont{Sawicki}},
  \bibinfo{journal}{Phys.\ Rev.} \textbf{\bibinfo{volume}{D75}},
  \bibinfo{pages}{044004} (\bibinfo{year}{2007}).

\bibitem[{\citenamefont{Eling and Jacobson}(2006{\natexlab{a}})}]{Aethersph}
\bibinfo{author}{\bibfnamefont{C.}~\bibnamefont{Eling}} \bibnamefont{and}
  \bibinfo{author}{\bibfnamefont{T.}~\bibnamefont{Jacobson}},
  \bibinfo{journal}{Class.\ Quant.\ Grav} \textbf{\bibinfo{volume}{23}},
  \bibinfo{pages}{5625} (\bibinfo{year}{2006}{\natexlab{a}}).

\bibitem[{\citenamefont{Eling and Jacobson}(2006{\natexlab{b}})}]{AetherBH}
\bibinfo{author}{\bibfnamefont{C.}~\bibnamefont{Eling}} \bibnamefont{and}
  \bibinfo{author}{\bibfnamefont{T.}~\bibnamefont{Jacobson}},
  \bibinfo{journal}{Class.\ Quant.\ Grav} \textbf{\bibinfo{volume}{23}},
  \bibinfo{pages}{5643} (\bibinfo{year}{2006}{\natexlab{b}}).

\bibitem[{\citenamefont{Kosteleck\'{y} and Samuel}(1989)}]{Kost1}
\bibinfo{author}{\bibfnamefont{V.~A.} \bibnamefont{Kosteleck\'{y}}}
  \bibnamefont{and} \bibinfo{author}{\bibfnamefont{S.}~\bibnamefont{Samuel}},
  \bibinfo{journal}{Phys.\ Rev.} \textbf{\bibinfo{volume}{D40}},
  \bibinfo{pages}{1886} (\bibinfo{year}{1989}).

\bibitem[{\citenamefont{Clayton}(2001)}]{Clayton}
\bibinfo{author}{\bibfnamefont{M.~A.} \bibnamefont{Clayton}}
  (\bibinfo{year}{2001}), \bibinfo{note}{preprint: \ttfamily
  gr-qc/0104103\rmfamily}.

\bibitem[{\citenamefont{Giannios}(2005)}]{Tevessph}
\bibinfo{author}{\bibfnamefont{D.}~\bibnamefont{Giannios}},
  \bibinfo{journal}{Phys.\ Rev.} \textbf{\bibinfo{volume}{D71}},
  \bibinfo{pages}{103511} (\bibinfo{year}{2005}).

\bibitem[{\citenamefont{Bekenstein}(2006)}]{BekReview}
\bibinfo{author}{\bibfnamefont{J.~D.} \bibnamefont{Bekenstein}},
  \bibinfo{journal}{Contemp.\ Phys.} \textbf{\bibinfo{volume}{47}},
  \bibinfo{pages}{387} (\bibinfo{year}{2006}).

\bibitem[{\citenamefont{Skordis et~al.}(2006)\citenamefont{Skordis, Mota,
  Ferreira, and B{\oe}hm}}]{Skor}
\bibinfo{author}{\bibfnamefont{C.}~\bibnamefont{Skordis}},
  \bibinfo{author}{\bibfnamefont{D.~F.} \bibnamefont{Mota}},
  \bibinfo{author}{\bibfnamefont{P.~G.} \bibnamefont{Ferreira}},
  \bibnamefont{and} \bibinfo{author}{\bibfnamefont{C.}~\bibnamefont{B{\oe}hm}},
  \bibinfo{journal}{Phys.\ Rev.\ Letters} \textbf{\bibinfo{volume}{96}},
  \bibinfo{pages}{011301} (\bibinfo{year}{2006}).

\bibitem[{\citenamefont{Will}(1993)}]{Will}
\bibinfo{author}{\bibfnamefont{C.~M.} \bibnamefont{Will}},
  \emph{\bibinfo{title}{Theory and Experiment in Gravitational Physics}}
  (\bibinfo{publisher}{Cambridge University Press}, \bibinfo{year}{1993}),
  chap.~\bibinfo{chapter}{5}, pp. \bibinfo{pages}{140--1},
  \bibinfo{edition}{revised edition} ed.

\bibitem[{\citenamefont{Dodelson and Liguori}(2006)}]{TeVeScosmo1}
\bibinfo{author}{\bibfnamefont{S.}~\bibnamefont{Dodelson}} \bibnamefont{and}
  \bibinfo{author}{\bibfnamefont{M.}~\bibnamefont{Liguori}},
  \bibinfo{journal}{Phys.\ Rev.\ Lett.} \textbf{\bibinfo{volume}{97}},
  \bibinfo{pages}{231301} (\bibinfo{year}{2006}).

\end{thebibliography}

\end{document}